\documentclass[preprint,aps,superscriptaddress,nofootinbib,showpacs]{revtex4}
\usepackage{graphicx}
\usepackage{epsfig}
\usepackage{bm}
\usepackage{amssymb}
\usepackage{wrapfig}

\newcommand{\asla}{\ooalign{\hfil/\hfil\crcr{$a$}}}

\newcommand{\psla}{\ooalign{\hfil/\hfil\crcr{p}}}
\newcommand{\ksla}{\ooalign{\hfil/\hfil\crcr{k}}}

\newcommand{\el}{{\cal L}}

\newcommand{\cA}{{\cal A}}
\newcommand{\cB}{{\cal B}}
\newcommand{\cC}{{\cal C}}

\newcommand{\psibar}{\mbox{$\overline{\psi}$}}

\newcommand{\tldW}{\mbox{${\tilde W}$}}

\newcommand{\epsi}{\mbox{$\varepsilon$}}

\newcommand{\va}{\mbox{$\bm{a}$}}
\newcommand{\vp}{\mbox{$\bm{p}$}}
\newcommand{\vq}{\mbox{$\bm{q}$}}
\newcommand{\vbr}{\mbox{$\bm{r}$}}
\newcommand{\vk}{\mbox{$\bm{k}$}}
\newcommand{\vn}{\mbox{$\bm{n}$}}

\newcommand{\vB}{\mbox{$\bm{B}$}}

\newcommand{\vR}{\mbox{$\bm{R}$}}

\newcommand{\vgamma}{\mbox{$\bm{\gamma}$}}
\newcommand{\vsigma}{\mbox{$\bm{\sigma}$}}

\newcommand{\sigm}{\mbox{$\langle \sigma \rangle$}}
\newcommand{\psiN}{\mbox{$\psi_N$}}
\newcommand{\psibN}{\mbox{${\bar{\psi}}_N$}}
\newcommand{\psiL}{\mbox{$\psi_\Lambda$}}
\newcommand{\psibL}{\mbox{${\bar \psi}_\Lambda$}}
\newcommand{\gs}{\mbox{$g_\sigma$}}
\newcommand{\gv}{\mbox{$g_\omega$}}

\newcommand{\mvq}{\mbox{$m_\omega^2$}}
\newcommand{\gsL}{\mbox{$g_\sigma^\Lambda$}}
\newcommand{\gvL}{\mbox{$g_\omega^\Lambda$}}

\begin{document}

\title{Relativistic Mean-Field Treatment of Pulsar Kick 
from Neutrino Propagation in Magnetized Proto-Neutron}


\author{Tomoyuki~Maruyama}
\affiliation{College of Bioresource Sciences,
Nihon University,
Fujisawa 252-8510, Japan}
\affiliation{Advanced Science Research Center,
Japan Atomic Energy Research Institute, Tokai 319-1195, Japan}
\affiliation{National Astronomical Observatory of Japan, 2-21-1 Osawa, Mitaka, Tokyo 181-8588, Japan}

\author{Nobutoshi~Yasutake}
\affiliation{Research Institute for the Early Universe,
University of Tokyo, Hongo 7-3-1, Bunkyo-ku, Tokyo 113-0033, Japan}

\author{Myung-Ki Cheoun}
\affiliation{Department of Physics, Soongsil University, Seoul,
156-743, Korea}
\affiliation{National Astronomical Observatory of Japan, 2-21-1 Osawa, Mitaka, Tokyo 181-8588, Japan}

\author{Jun~Hidaka}
\affiliation{National Astronomical Observatory of Japan, 2-21-1 Osawa, Mitaka, Tokyo 181-8588, Japan}

\author{Toshitaka~Kajino}
\affiliation{National Astronomical Observatory of Japan, 2-21-1 Osawa, Mitaka, Tokyo 181-8588, Japan}
\affiliation{Department of Astronomy, Graduate School of Science, University of Tokyo, Hongo 7-3-1, Bunkyo-ku, Tokyo 113-0033, Japan}

\author{Grant J. Mathews}
\affiliation{Center of Astrophysics, Department of Physics,
University of Notre Dame, Notre Dame, IN 46556, USA}

\author{Chung-Yeol Ryu}
\affiliation{Department of Physics, Soongsil University, Seoul, 156-743, Korea}

\date{\today}


\begin{abstract}
We make a perturbative calculation of  neutrino scattering and absorption
in hot and dense hyperonic neutron-star matter in the presence
of a strong magnetic field.
We calculate  that the absorption cross-sections 
in a fully relativistic mean-field theory.    
We find that there is a remarkable angular dependence, 
{\it i.e.} the neutrino absorption strength is reduced in a direction parallel
to the magnetic field and enhanced in the opposite direction.
This asymmetry in the neutrino absorption is estimated to be as much as  
2.2 \% of the entire neutrino momentum for an interior magnetic field 
of $\sim 2 \times 10^{17}$G.
The pulsar kick velocities associated with this asymmetry  are shown 
to be comparable to  observed velocities.
\end{abstract}

\maketitle


\newpage

\section{Introduction}

Hot and dense hadronic matter is a topic of considerable current interest
in nuclear and particle physics as well as  astrophysics
because of its associated exotic phenomena.
In particular, many studies  have
addressed the possible exotic phases of  high density
matter. Neutron stars are  thought to be the most realistic possible sites to study the physics of high density matter.
For example, the possible existence of an anti-kaon condensation in
neutron stars has been suggested  \cite{kn86}, and the possible implications for its astrophysical
phenomena have been widely discussed \cite{lbm95,t95,l96,pb97}.

These discussions, however,  heavily depend upon the nuclear-matter 
equation of state (EOS), which governs both the static and dynamic
properties of neutron stars.  
Hence, many papers \cite{tpl94,K-con,fmmt96,g01,bb94,p00,ty99,ny07} have
been devoted to the study of  the neutron-star EOS.
In particular, the thermal evolution of neutron stars
by neutrino emission is a topic of considerable interest 
\cite{bkpp88,t88,fmtt94,pb90,t98,yakolov01,ny09} 
regarding the  dynamical evolution  of neutron stars.
For example, Reddy {\it et al.} \cite{rml98} studied neutrino
propagation in proto-neutron stars (PNSs) as a means to examine 
the hyperon phase in the high density region.

On the other hand, since the discovery of  magnetars \cite{pac92,mag3}, magnetic fields are thought to play an
important role in many astrophysical phenomena such as the development of asymmetry in
supernova (SN) remnants. Indeed, strong
magnetic fields turn out to be a crucial ingredient for the still poorly understood
mechanism to produce  non-spherical SN explosions, pulsar kicks~\cite{lyne94}, i.e. the
 high  velocity~\cite{rothchild94} that some PNSs receive at birth.

Although several post-collapse instabilities have been  studied as a
possible source of non-spherical explosions and  pulsar kicks, the unknown origin of the
initial asymmetric perturbations and the uncertainties  in
the numerical simulations make this possibility
difficult to unambiguously verify~\cite{burrows06,marek09}.
Another viable candidate is the possibility of
asymmetric neutrino emission either as a result of  parity violation
in the weak interaction~\cite{vilenkin95,horowitz98} or as a result of
an asymmetric magnetic field~\cite{bisnovat93} in strongly
magnetized PNSs.

In this work, we take the asymmetric neutrino emission as one of the main reasons for the asymmetric phenomena observed in the PNS.
This asymmetric neutrino emission is assumed  be caused by the two processes;
one is the asymmetric production inside PNSs;
and the other is the damping of the neutrino luminosity
through  neutrino absorption in the nuclear medium.

The direct and modified URCA processes may play a role in  neutrino emission,
but the main effect of these processes is in the neutron-star cooling  \cite{sawyer79,yakolov01}, where
in-medium effects play an important role \cite{frimax79,BRSSV95}.
Of course, a strong magnetic field leads to an angular-dependence of the neutrino production
in the URCA process
because of the spin polarization of electrons and positrons in matter
\cite{chugai84,dorofeev}.  Nevertheless,  we assume here that the URCA process is not  important in the PNS stage.

Other  effects, such as the Landau levels due to the magnetic field \cite{kisslinger2,kisslinger3}, the angular dependence of the neutrino production
caused by a possible  pion condensation phase \cite{vosk86,parenkov}, and a possible quark-matter
color-super conducting phase \cite{berderman} {\it etc} are also assumed to be small  in this work.


Over a decade ago, Lai {\it et al.}~\cite{arras99,lai98} calculated
 the neutrino-nucleon scattering during  neutrino propagation inside a neutron star in the context of  a non-relativistic framework~\cite{arras99}.  Within that approximation they  showed
 that even a
$\sim$1\% asymmetry in the total neutrino
luminosity of $\sim 10^{53}$ ergs could be enough to explain the
observed pulsar kick velocities.

Kusenko, Segre and Vilenkin \cite{KSV98} criticized this conclusion and
theoretically showed that the asymmetry in the neutrino scattering cross-section
does not lead to an asymmetry in the neutrino emission if the system is in
complete thermodynamic equilibrium.
However, they only considered only neutrino-neutron collisions and
neglected the Fermi-Dirac statistics.
Hence, their proof is only applicable in the very low-density region.
Furthermore,  neutrino scattering inside dense nuclear matter does not play a role in
either the thermal evolution or the propagation of the
non-equilibrium part of the neutrinos.
On the other hand, the absorption part of the collisions may make a large
contribution  to the asymmetry \cite{MKYCR11}.  That is what we demonstrate here.

On the other hand, the past decades have seen many successes in the relativistic
treatment of the nuclear many-body problem.
The relativistic framework has several advantages \cite{SW86,serot97}.  Among them this formalism provides
 a useful Dirac phenomenology for the
description of nucleon-nucleus scattering \cite{Hama,Tjon},
a natural means to incorporate  the spin-orbit force \cite{serot97}, 
and a reliable means to compute the structure of extreme nuclei \cite{hirata}.
These results have shown that there are large attractive scalar
and repulsive vector fields, and that the nucleon
effective mass becomes small in the nuclear medium.
This mechanism may drive the self-suppression mechanism of 
kaon-condensation in in nuclear matter, and may lead to a stable kaon condensation phase  in neutron stars (NSs)
\cite{K-con}.

In Ref.~\cite{MKYCR11}
we reported results for the first time on the neutrino
absorption cross-sections in hot dense
magnetized NS matter calculated in a fully
relativistic mean field (RMF) theory~\cite{SW86,serot97} including hyperons.
In that work  we took into account the Fermi motion of
baryons and electrons, their recoil effects, distortion effects of the
Fermi spheres by the magnetic field, and effects from the
energy difference of the mean field between initial and final
baryons in a fully relativistic framework.
We found that even a few percent breaking of  isotropic symmetry in the
neutrino absorption cross-section may cause an asymmetric
emissions of neutrinos from PNSs.

In this paper, we provide more detailed explanations of the
neutrino scattering and absorption cross-sections in
magnetized NS matter in the context of RMF theory. 
We then solve the Boltzmann equation for neutrino transport
in a 1D model and discuss implications of our numerical results for
 pulsar kicks. 
In particular, we focus on the collision between a neutrino and a
particle in nuclear matter in the presence of a strong magnetic field
and a core temperature of $20 - 40$MeV. 
Two-baryon process are not taken into account in the present PNS
calculation since they only play an important role
at low temperature ($\sim$ a few MeV) \cite{sawyer79}.

In Sec. II we introduce our EOS for nuclear matter based upon the RMF theory.
In Sec. III we explain the neutrino scattering and absorption cross-sections
in  baryonic matter in the presence of  strong magnetic fields.
Numerical results and detailed discussions of neutrino reactions and
propagation in baryonic matter at finite temperature are presented in Sec. IV. Summaries are given in Sec. V
with further arguments on the associated pulsar kicks of magnetized PNSs.
Finally, in Sec. VI,  as topics for  future work, we discuss  other plausible characteristics of PNS interiors that  may affect the pulsar kicks.

\newpage

\section{Neutron-Star Matter in the Relativistic Mean-Field Approach}

In this work  we calculate  neutrino cross-sections in
neutron-star matter in the RMF approach.
For this purpose we define the Lagrangian density as 
\begin{equation}
\el  =  \el_{Lep} +  \el_{RMF} +  \el_{Mag} +  \el_{W} ,
\end{equation}
where the first, second, third and fourth  terms are the lepton, RMF,
 magnetic, and weak interaction parts, respectively.  We consider NS matter including nucleons, Lambdas,
electrons and electro-neutrinos $(\nu_e)$.
 Detailed expressions for the magnetic and weak parts are
explained in the next section.


The lepton and RMF parts of the Lagrangian density utilized in this work are given as
\begin{eqnarray}
\el_{Lep} &=& \psibar_{\nu}  i{\gamma_\mu \partial^\mu}\psi_{\nu}
+ \psibar_{e}  (i{\gamma_\mu \partial^\mu} - m_e )\psi_{e} ,
\\
\el_{RMF} & = &
{\psibN} (i{\gamma_\mu \partial^\mu} - M_N )\psiN
+ \gs {\psibN} \psiN \sigma
+ \gv {\psibN} \gamma_\mu \psiN \omega^\mu
\nonumber \\
&&+ \psibL (i{\gamma_\mu \partial^\mu} - M_{\Lambda} )\psiL
+ \gsL \psibL \psiL  \sigma
+ \gvL \psibL \gamma_\mu \psiL \omega^\mu
\nonumber \\
&& - {\widetilde U} [\sigma]
+ \frac{1}{2} m_{\omega}^2 \omega_\mu \omega^\mu
-\frac{C_{IV}}{2M_N^2} {\psibN} \gamma_\mu \tau_a \psi_N
{\psibN} \gamma^\mu \tau_a \psi_N ,
\label{RMFLag}
\end{eqnarray}
where $\psi_\nu$, $\psi_e$, $\psiN$, $\psiL$, $\sigma$, and $\omega$ are
the electron neutrino,  and electron, nucleon,
Lambda, sigma-meson and omega-meson  fields, respectively, with corresponding masses
$m_e$, $M_N$, $M_{\Lambda}$, and $m_{\omega}$.
${\widetilde U} [\sigma]$ is the self-energy potential of the
scalar mean-field given in Refs.~\cite{K-con,TOMO1}.
The last term describes the vector isovector interaction between two
nucleons, which is equivalent to $\rho$-meson exchange \cite{SW86}. We
adopt natural units, {\it i.e.}~$\hbar = c = 1$.

From the Euler-Lagrange equation of the above Lagrangian, the
Dirac spinor of the baryon $u_b(\vp,s)$ is obtained as a solution to the following
equation
\begin{eqnarray}
\left[ \psla - M_b^* - U_0 (b) \gamma_0 \right] u_b(\vp,s) = 0 ,
\end{eqnarray}
where $U_0(b)$ is the time component
of the  mean-field vector potential.
We hereafter introduce the Feynman dagger $\psla \equiv \gamma_\mu p^\mu$
for convenience. The baryon effective masses $M_b^*$ are given by
\begin{eqnarray}
M_N^* & = & M_N - U_{s}(N),~\nonumber \\
M_\Lambda^{*} & = &  M_\Lambda - U_{s}({\Lambda}),
\label{efmas}
\end{eqnarray}
with the scalar mean-field potentials
\begin{eqnarray}
U_{s}(N) =   \gs \sigm , &~~~&
U_{s}(\Lambda) =  \gsL \sigm.
\end{eqnarray}
The scalar mean-field $\sigm$ is given by
\begin{equation}
\frac{\partial}{\partial \sigm} {\tilde U} [\sigm]
= \gs \left[ \rho_s(p) + \rho_s(n) \right] + \gsL \rho_s(\Lambda) ,
\end{equation}
with the scalar densities
\begin{equation}
\rho_s (b) \equiv \frac{2}{(2 \pi)^3} \int {\rm d}^3 {\vp} ~
\left[ n^{(+)}_b [e_b^{(+)}({\vp})]  + n^{(-)}_b [e_b^{(-)}(\vp) ] \right]
\frac{M_{b}^*}{E^*_b(\vp)}~.
\label{rhos}
\end{equation}
Here, $e_b^{(\pm)}$ are the single particle $(+)$ and antiparticle ($-$)
energies,
$E^*_b(\vp) = \sqrt{ {\vp}^2 + M_b^{*2} }$,
and the Fermi distributions, $n^{(\pm)}_b(e_b^{(\pm)})$, are defined as usual,
\begin{equation}
n^{(\pm)}_b (e_b^{(\pm)}) = \frac{1}{1 + \exp[(e_b^{(\pm)} \pm \epsi_b)/T]}~,
\end{equation}
in terms of  the temperature $T$ and the chemical potential $\epsi_b$.

In addition, the baryon single-particle energies are  written as
$e_{b}^{(\pm) }({\vp}) =  E_b^* (\vp) \pm U_0(b)$,
with  the $U_0(b)$  calculated as
\begin{eqnarray}
U_{0}(p) & = & \frac{\gv}{\mvq} \left\{ \gv (\rho_p + \rho_n )
+ \gvL \rho_{\Lambda} \right\}
+ \frac{C_{IV}}{M_N^2} (\rho_p - \rho_n ) ,
 \\
U_{0}(n) & = & \frac{\gv}{\mvq} \left\{ \gv (\rho_p + \rho_n )
+ \gvL \rho_{\Lambda} \right\}
- \frac{C_{IV}}{M_N^2} (\rho_p - \rho_n ) ,
 \\
U_{0}(\Lambda) & = & \frac{\gvL}{\mvq} \left\{ \gv (\rho_p + \rho_n )
+ \gvL \rho_{\Lambda} \right\}
\end{eqnarray}
in terms of  the proton, neutron and Lambda number densities,
$\rho_p$, $\rho_n$ and $\rho_\Lambda$.

In this work, neutron-star matter at finite temperature
 includes  protons, neutrons, Lambdas($\Lambda$s), electrons and neutrinos.  These  are constrained by the conditions of charge neutrality and  beta equilibrium.
Therefore, the proton number density is equal to the electron number density,
$\rho_p = \rho_e$, and the chemical potentials obey the following condition
\begin{equation}
\epsi_n  = \epsi_{\Lambda} = \epsi_p + \epsi_e~.
\end{equation}
The lepton fraction is also fixed as $Y_L = (\rho_e + \rho_\nu)/\rho_B$ with $\rho_B = \rho_p + \rho_n + \rho_\Lambda$.
\begin{figure}[ht]
\begin{center}
{\includegraphics[scale=0.55,angle=270]{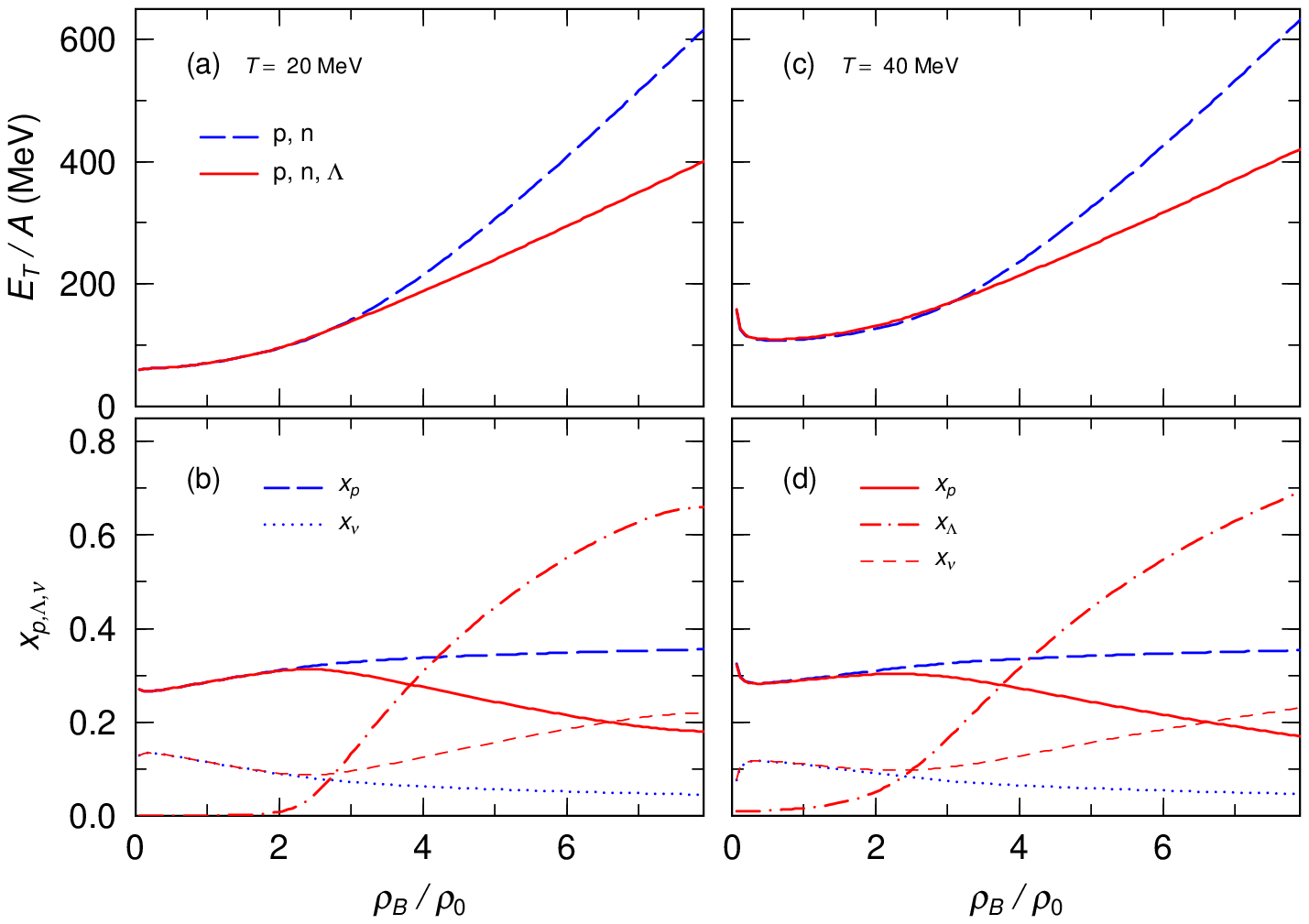}} \caption{\small
(Color online) Upper panels (a) and (c) show the density dependence
of the total energy per baryon $E_T/A$ in  neutron-star matter
for $T$ = 20 MeV (a) and 40 MeV (c).
Solid and long-dashed lines represent results with and without $\Lambda$ particles.
Lower panels (b) and (d) show number fractions of protons $x_p$,
$\Lambda$ particles $x_\Lambda$, and neutrinos $x_\nu$ for $T$ = 20 MeV
(b) and 40 MeV (d).
Solid, dot-long-dashed, and short-dashed lines represent proton, Lambda, and neutrino number fractions,
respectively.
Long-dashed and dotted lines represent the calculated
proton and neutrino number fractions in a system without $\Lambda$s.
In the present calculations we use the parameter-set PM1-L1 \cite{kpcon}
for the RMF and the lepton fraction is set to $Y_L = 0.4$. }
\label{EOSf}
\end{center}
\end{figure}

Since we focus only on the asymmetry of neutrino emission
caused by the presence of a magnetic field and the
existence of  strange matter, we choose one parameter-set, PM1-L1
\cite{kpcon}, in order not to distract the discussion. 
This parameter set gives the binding energy per baryon $BE = 16$ MeV,
a nucleon effective mass of $M_N^*/M_N=0.7$ and an incompressibility 
parameter of $K=200$MeV at $\rho_0 $= 0.17 fm$^{-3}$ in nuclear matter.  
The sigma- and omega-Lambda couplings are 2/3 of those
for the nucleon, $g_{\sigma, \omega}^\Lambda = \frac{2}{3} g_{\sigma,
\omega}$. 
Similar relations are used in the quark
meson coupling (QMC) model \cite{tsht98}.

In Fig.~\ref{EOSf} we show the energy per nucleon, which is a kind of 
the equation of state (EOS), in the upper
panels (a and c) and the proton and Lambda fractions in the lower
panels (b and d) at $T=20$MeV (a and b) and $T=40$MeV (c and d). In these
calculations the lepton fraction is taken to be $Y_L = 0.4$.
Solid and dashed lines represent the results for matter with
and without $\Lambda$s, respectively.
Dot-dashed lines
in the lower panels indicate the Lambda fraction, which appears when  $\rho_B \gtrsim 2 \rho_0$ and significantly affects
 the EOS for $\rho_B \gtrsim 3 \rho_0$.

Here we should comment about the anti-particle contribution.
The  density of  anti-neutrinos is less than 0.5 \%
of the neutrino density when $\rho_B = \rho_0$ and $T=40$MeV. This  ratio is much lower than other particles.
With  larger density and lower temperature, this ratio becomes smaller.
Thus,  anti-particles does not significantly  contribute to the EOS or other
observables as  discussed below.

When Lambda particles are not included, the PM1-L1 EOS is sufficiently 
stiff \cite{K-con} to allow a neutron star with mass larger 
than the value observed observed for PSR J1614-2230 
of $M = 1.97 \pm 0.04 M_\odot$  \cite{T-Solar}.
When the Lambda particles are  included, however, the EOS becomes softer and does not allow such a large maximum
neutron-star mass.
This could be resolved if  we introduce additional repulsive force between $\Lambda$s \cite{Bednarek11} consistent with hypernuclear data.  
Another possibility would be introducing a repulsive three-body force.

In this paper, however,  our goal is to explore the effects of magnetic 
fields in generating pulsar kicks and  not to discuss the ambiguity of 
the mean-field EOS in regards to the maximum neutron-star mass.  
In this work, therefore, only the $\Lambda$ particle is introduced as a 
hyperon. 
One could also introduce a Sigma ($\Sigma$) mean-field in matter, 
which is repulsive \cite{noumi02} and appears at rather high density. 
However,  its abundance fraction is small \cite{ryu07,ryu12}.
Though the Xi ($\Xi$) particle may be attractive \cite{KV68}, we 
we do not have sufficient information about the Xi ($\Xi$) particle and ignore
its contribution.

\newpage

\section{Cross-Sections for Neutrino Reactions in Magnetized Proto-Neutron Star Matter}

\subsection{Dirac Wave Function in a Magnetic Field}

We assume a uniform magnetic field along
the $z$-direction $\vB = B {\hat z}$ with $B ^<_\sim 10^{18}$ G.  For this field strength
the effect of the magnetic field on baryons is small enough
to be treated perturbatively.
The magnetic part of the Lagrangian density is written as
\begin{equation}
\el_{Mag} =  \el_{BM} + \el_{eM},
\label{magL}
\end{equation}
where the first and second terms describe the magnetic interactions of
baryons and electrons, respectively.

Considering
only the spin-interaction term, the baryon magnetic-interaction Lagrangian
density can be written as
\begin{equation}
\el_{BM} = \sum_b \mu_b {\bar \psi}_b \sigma_{\mu \nu} \psi_b
F^{\mu \nu} = \sum_b \mu_b {\bar \psi}_b \sigma_z \psi_b B
\end{equation}
with the electromagnetic tensor given by
$F^{\mu \nu} =\partial^\mu A^\nu - \partial^\nu A^\mu$,
where $A^\mu$ is the electromagnetic vector potential,
$\sigma_{\mu \nu} = [\gamma_\mu, \gamma_\nu]/2i$,
$\sigma_z = {\rm diag}(1,-1,1,-1)$ and $\mu_b$ is the baryon magnetic moment.
The baryon wave functions can be obtained by solving the following Dirac equation
\begin{equation}
\left[ \psla - M^*_b -  U_0 (b) \gamma_0 - \mu_b B
\gamma_0\sigma_z \right ] u_b(p,s) =0 ~. \label{DiracE}
\end{equation}
The single particle energies $e_b (\vp,s)$ and the Dirac spinors in the limit
of a weak magnetic field are given as
\begin{eqnarray}
e_b (\vp,s) &=&
\sqrt{ p_z^2 + \left( \sqrt{\vp_T^2 + M^{*2}_b} + \mu_b B s\right)^2 }
+ U_0(b)
\nonumber \\
& \approx &
E_b^*(\vp) + U_0(b) + \Delta E_b^*(\vp) s ~~
\label{sgEn}
\end{eqnarray}
with
\begin{equation}
\Delta E_b^*(\vp) = \frac{\sqrt{\vp_T^2 + M^{*2}}}{E^*_b(\vp)} \mu_b B ~~,
\label{DeltaE}
\end{equation}
and
\begin{eqnarray}
u_b(\vp,s) {\bar u}_b(\vp,s) &=&
\frac{1}{4 E_b^*(\vp)}[E_b^*(\vp) \gamma_0 - \vp \cdot {\vgamma} + M^*_{b}]
(1 + s \gamma_5 \asla(\vp))
\nonumber \\ &&  ~~
+  \frac{p_z \mu_b B}{ 4 E^{*3}_b(\vp)}
\left( {\vsigma} \cdot \vp - M_b^* \gamma_5 \gamma_0 \right)
\nonumber \\ && ~~
+ \frac{s \mu_b B}{ 8 E_b^*(\vp) \sqrt{\vp_T^2 + M_b^{*2}} }
\left(- E_b^*(\vp) \gamma^0 + M^*_{b}
+ p_z \gamma^3 - p_x \gamma^1 - p_y \gamma^2
\right) ,
\label{uubarB}
\end{eqnarray}
with
\begin{equation}
a (\vp) \equiv (a_0, \va_T, a_z) =
\frac{1}{\sqrt{\vp_T^2 + M_b^{*2}}} (p_z,0,0,E_b^*(\vp))~~ .
\end{equation}
Detailed derivations of these expressions of Eq.~(\ref{uubarB}) 
are presented in appendix \ref{DSapp}.
The second and third terms of Eq.~(\ref{uubarB})
do not appear in the non-relativistic framework, but their
contributions are negligibly small and can be omitted in the present work.

For the electron contribution in Eq.~(\ref{magL}),
we have to use another treatment. This is  because electron mass is very small,
and its current is almost a Dirac current
\begin{equation}
\el_{eM} = - e \psibar_e \gamma_\mu \psi_e A^\mu,
\end{equation}
where $\psi_e$ is the electron field.
Also, the effect of a strong magnetic field on electrons
may not be a small perturbation.
The electron energy in the presence of a  strong magnetic
field is generally given by
\begin{eqnarray}
e_e(n,k_z;s)&=& \sqrt{k_z^2 + m_e^2 + eB(2n + 1 - s)},
\end{eqnarray}
where $n$ stands for the Landau levels of the electrons.

But the electron wave function
also becomes a plane wave in the limit of $B \rightarrow 0$, so that we can use the same expression as Eq. (\ref{uubarB}) for electrons,
aside from the spin vector.
The upper component of the electron Dirac spinor is an eigenvector
of the matrix $\sigma_z$.  The spin vector in the rest
frame of the electron is then $(0;0,0,1)$.
In the matter frame the boosted spin vector can be  written as
\begin{equation}
a(\vk) = a_e(\vk)
\equiv \left( \frac{k_z}{m_e},  \frac{k_z \vk_T}{m_e (E_e(\vk) + m_e)},
1 + \frac{k_z^2}{m_e (E_e(\vk) + m_e)} \right) ~~,
\end{equation}
where $k_z$ and $\vk_T$ are the components along  the $z$-direction
 and  perpendicular to the $z$-direction, respectively.

When $\sqrt{2e B} \ll \epsi_e$,  the summation over $n$
can be approximated as an  integration over energy, i.e.
\begin{equation}
\sum_n ~~ \rightarrow ~~
\frac{1}{2 eB} \int d x_T  ~~ , ~~~~
(x_T = 2e B(n + \frac{1}{2} ) ) ~~.
\end{equation}
Note that the variable $x_T$ corresponds to $\vk_T^2$ in the limit
of $B \rightarrow 0$.
Then, the expectation value of an operator $\hat{\cal O}$ is given by
\begin{eqnarray}
&&<\hat{\cal O}> =
\frac{2 e B}{(2 \pi)^2} \sum_s \sum_n \int d k_z
n_e \left( e_e (n,k_z,s) \right) {\cal O}(n,k_z,s)
\nonumber \\ &\approx&
\frac{1}{(2 \pi)^2} \sum_s \int d x_T \int d k_z
n_e\left( e_e(x_T,k_z,s) \right) {\cal O} (x_T,k_z,s)
\nonumber \\ &\approx&
\frac{1}{(2 \pi)^3} \sum_s \int d^3 k
n_e \left( e_e(\vk,s) \right) {\cal O}(\vk,s) ,
\end{eqnarray}
where the electron energy is approximately given as $e_e =  \sqrt{\vk^2 + m_e^2 - eBs}$.

Actual calculations are performed in the limit of $m_e
\rightarrow 0$, so that  the electron energy and the spin vector are approximated by
\begin{eqnarray}
&&e_e \approx \sqrt{\vk^2 + m_e^2} - \frac{eBs}{2 \sqrt{\vk^2 + m_e^2}}
\approx |\vk| + \frac{m_e}{|\vk|} \mu_e B s~~ ,
\\
&&
a_e(\vk) \approx \frac{1}{m_e}
\left( k_z,~ \frac{k_z \vk_T}{|\vk|},~ \frac{k_z^2}{|\vk|} \right) ,
\end{eqnarray}
where $\mu_e = -e /2 m_e$~~.

As already commented at the end of Sec. III, the fractions of the anti-leptons and anti-baryons are
negligibly small, and these particles do not contribute to the neutrino reactions.
Therefore, we ignore the contributions
from antiparticles, and omit the superscript '+' in the single
particle energies $e_b^{(\pm)}(\vp)$ and the Fermi distribution
$n_b^{(\pm)}(\vp,s)$.

\subsection{Neutrino Reaction Cross-Sections}

In this subsection we consider neutrino reactions in NS matter consisting
of electrons and baryons (i.e. protons, neutrons and $\Lambda$s).
The weak interaction part of the Lagrangian density ${\cal L}_W$ in Eq.~(1) is written as
\begin{eqnarray}
\el_{W} &=& \frac{G_F}{2} \left\{\sum_{\alpha,\beta}
\psibar_\alpha \gamma_\mu (c_V-c_A\gamma_5) \psi_\beta \right\}^2,
\label{elLag}
\end{eqnarray}
where the indices $\alpha$ and $\beta$ indicate particles
comprising the NS matter.
The $c_V$ and $c_A$ are the weak vector and axial coupling constants dependent on each channel.

We utilize the  impulse approximation, {\it i.e.} individual collisions between the initial neutrino
and the constituent particles.  We consider both  neutrino scattering ($\nu_e \rightarrow \nu_e'$) channels
\begin{eqnarray}
\nu_e  + p~ &\rightarrow& \nu_e' + p', \\
\nu_e  + n~ &\rightarrow& \nu_e' + n', \\
\nu_e  + \Lambda~ &\rightarrow& \nu_e' + \Lambda', \\
\nu_e  + e^{-} &\rightarrow& \nu_e' + e'^-,
\end{eqnarray}
and absorption ($\nu_e \rightarrow e^{-}$) channels
\begin{eqnarray}
\nu_e  + n &\rightarrow& e^{-} + p~, \\
\nu_e  + \Lambda &\rightarrow& e^{-} + p~~.
\end{eqnarray}

As noted above,  we consider rather low temperatures, $T \ll \epsi_b$.  
Therefore, we may ignore the contribution from antiparticles.
In addition, we treat this system as partially spin-polarized
 owing to the magnetic field.
The cross-section can then be described in terms of  the initial and
final lepton momenta $\vk_i$ and $\vk_f$
\begin{eqnarray}
\frac{d^3 \sigma}{d k_f^3} &=&
\frac{G_F^2}{16 \pi^2} V \sum_{\alpha,\beta}
\sum_{s_{l}, s_{i}, s_{f}} [1- n_{l} ( e_{l}(\vk_f,s_l) )]
\int  \frac{d^3 p_i}{(2 \pi)^3} \frac{d^3 p_f}{(2 \pi)^3}
W_{BL} (k_i,k_f,p_i,p_f; \alpha,\beta)
\nonumber \\ && ~~~~~~ \times
n_{\alpha} ( e_{\alpha}(\vp_i,s_{i}) )[1- n_{\beta}(e_{\beta} (\vp_f, s_f))]
\nonumber \\ && ~~~~~~ \times
(2 \pi)^4 \delta^3 (\vk_i + \vp_i - \vk_f - \vp_{f})
\delta ( |\vk_i| + e_{\alpha} (\vp_i) - e_{\beta} (\vp_f) - e_{l}(\vk_f))  ,
\label{CrsLB}
\end{eqnarray}
where $V$ is the volume of the system, and 
index $l$ denotes final lepton species.
Indices $\alpha$ and $\beta$ denote initial and final baryons and electrons, 
which have momenta $p_i$ and $p_f$, respectively.
The function $W_{BL}$ in Eq.~(\ref{CrsLB}) is defined as a product of 
lepton and hadron weak currents
\begin{equation}
W_{BL} = \frac{1}{4 |\vk_i| |\vk_f| E_{\alpha}^*(\vp_i)
E_{\beta}^*(\vp_f)}L^{\mu \nu} N_{\mu \nu}
\label{WBL}
\end{equation}
with
\begin{eqnarray}
L^{\mu \nu} &=& \frac{1}{4} {\rm Tr} \left\{ (\ksla_{f} + m_{l})(1
+ \gamma_5 \asla_{l} s_l )  \gamma^\mu (1 - \gamma_5) \ksla_{\alpha} \gamma^\nu
(1 - \gamma_5)
\right\}  ~~,
\end{eqnarray}
and
\begin{eqnarray}
N_{\mu \nu}&=&  \frac{1}{4} {\rm Tr}
\left\{ (\psla_f + M^*_{\beta})(1 + \gamma_5 \asla_{\beta} s_f)
\gamma_\mu (c_V - c_A \gamma_5)   \right.
\nonumber \\ && ~~~~~~~~~~ \times \left.
 (\psla_i + M^*_{\alpha})(1 + \gamma_5 \asla_{\alpha} s_{i})
\gamma_\nu (c_V  - c_A \gamma_5) \right\} ,
\end{eqnarray}
where $m_l$ is the mass of the final lepton.

Since we take the weak magnetic field limit, we treat this system 
as partially spin-polarized  owing to the magnetic field. 
Then the Fermi distribution and the delta function in the above
equations can be expanded in terms of the magnetic field $B$.
Finally, the cross-section can be summarized as a sum of two contributions, $\sigma^0_{S,A}$ independent of $B$ and $\Delta \sigma_{S,A}$ depending on $B$,
\begin{equation}
\frac{d^3 \sigma_{S,A}}{d k^3_f} =
\frac{d^3 \sigma_{S,A}^0}{d k^3_f} + \frac{d^3 \Delta \sigma_{S,A}}{d k^3_f},
\label{CrsNM}
\end{equation}
where the indices $S$ and $A$ indicate the cross-sections for
 scattering or  absorption, respectively.

For the absorption process, we use the energy delta function in 
Eq.~(\ref{CrsLB}) to further separate the magnetic part of 
the cross-section of Eq.~(\ref{CrsNM}) into two parts
\begin{equation}
\Delta {\sigma}_A = \Delta \sigma_M + \Delta \sigma_{el},
\label{Dsig}
\end{equation}
where first and second terms are the contribution from target particle 
and outgoing electron, which appear only in the absorption 
($\nu_e \rightarrow e^{-}$) process. 
Detailed derivations are written in the appendix \ref{NRCS}.

The first term of Eq.~(\ref{Dsig}), $\Delta \sigma_M$, is calculated as
\begin{equation}
\frac{d^2 \Delta \sigma_M}{d k_f d \Omega_f}
= \frac{4 \pi G_F^2 B}{(2 \pi)^6} \frac{|\vk_f|}{|\vk_i|}
(1 - n_{l} (|\vk_f|)) \sum_{\alpha,\beta} (T_A + T_B)~~,
\label{DsigM1}
\end{equation}
where
\begin{eqnarray}
T_A &=&
 \frac{1}{|\vq|}\int \frac{d^3 p_i}{ |\vp_i| E^{*}_{\alpha}}
\delta(t-t_p)
\left\{ n_{\alpha}^\prime (E_{\alpha}^*+U_0(\alpha)) [1 - n_{\beta}(E_{\beta}^* + U_0(\beta))] \mu_{\alpha} \tldW_{i}
\right. \nonumber \\ && \quad\quad\quad\quad \left. +~
n^\prime_{\beta} (E_{\beta}^{*} + U_0(\beta)) n_{\alpha} (E_{\alpha}^{*} + U_0(\alpha))
(\mu_{\alpha} \tldW_{i} - 2 \mu_{\beta} \tldW_f) \right\} ,
\nonumber \\
T_B& = & - \frac{1}{\vq^2}\int \frac{d^3 p_i}{\vp_i^2 E^*_{\alpha}}(E^*_{\alpha} + q_0)
\delta(t-t_p) n_{\alpha} (E_{\alpha}^* + U_0(\alpha))
\nonumber \\ && \quad \quad \quad \times
\left[ 1- n_{\beta} (E_{\beta}^* + U_0(\beta)) \right]
\left(\mu_{\alpha} \frac{\partial  \tldW_{i}}{\partial t}
- \mu_{\beta} \frac{\partial \tldW_f}{\partial t} \right)~~ ,
\label{tatb}
\end{eqnarray}
with
\begin{eqnarray}
\tldW_{i}
 &=&  c_V^2 \left\{ \left[k_f \cdot
(M^{*}_{\beta} p_i - M^*_i p_f) \right] (k_i \cdot b_{\alpha})
- \left[k_i \cdot (M^*_{\beta} p_i - M^*_{\alpha} p_f ) \right]
(k_f \cdot b_{\alpha}) \right\}
\nonumber \\ &&
+ c_A^2 \left\{\left[ - k_f \cdot (M^*_{\beta} p_i + M^*_{\alpha} p_f) \right]
(k_i \cdot b_{\alpha})
+ \left[k_i \cdot (M^*_{\beta} p_i + M^*_{\alpha} p_f) \right]
(k_f \cdot b_{\alpha}) \right\}
\nonumber \\ &&
- 2 c_V c_A M^*_{\alpha} \left\{ (k_f \cdot p_f)(k_i \cdot b_{\alpha})  +
(k_i \cdot p_f)(k_f \cdot b_{\alpha} ) \right\} ,
\\
\tldW_f
 &=&
 c_V^2 \left\{
\left[k_f \cdot (M_{\beta}^* p_i - M^*_{\alpha} p_f) \right] (k_i \cdot b_{\beta})
- \left[k_i \cdot (M^*_{\beta} p_i - M^*_{\alpha} p_f) \right]
( k_f \cdot b_{\beta} ) \right\}
\nonumber \\
&& + c_A^2 \left\{
\left[ k_f \cdot (M_{\beta}^* p_i + M_{\alpha}^* p_f) \right] (k_i \cdot b_{\beta})
- \left [k_i \cdot (M^*_{\beta} p_i + M^*_{\alpha} p_f ) \right] (k_f \cdot b_{\beta}) \right\}
\nonumber \\
&&  - 2 c_V c_A M^*_{\beta} \left\{ (k_i \cdot p_i)(k_f \cdot b_{\beta} )
+ (k_f \cdot p_i )(k_i \cdot b_{\beta} ) \right\}
\end{eqnarray}
and
\begin{equation}
b_{\alpha} =  \frac{\sqrt{\vp_T^2 + M_{\alpha}^{*2}}}{E_{\alpha}^*(\vp)}
a_{\alpha} (p_{\alpha}) ~~.
\end{equation}
In these equations the four momenta $p_i$ and $p_f$ are defined by
$p_i \equiv (E^*_{\alpha}(\vp_i), \vp_i)$ and $p_f \equiv (E^*_{\beta}(\vp_f), \vp_f)$.

When the target particle is an electron, the above expression is
slightly altered.
When both the initial and final particles are electrons,
the above equations are written as
\begin{eqnarray}
\tldW_i / m_e &=&  \delta_{\beta e}
\left\{ c_V^2 \left[ \left( k_f \cdot (p_i - p_f)\right) (k_i \cdot b_i)
- \left(k_i \cdot (p_i - p_f) \right) (k_f \cdot b_i) \right] \right.
\nonumber \\ &&  + c_A^2 \left[
\left(- k_f\cdot (p_i + p_f) \right) (k \cdot b_{\alpha})
+ \left(k_i \cdot (p_i +p_f)  \right) (k_f \cdot b_{\alpha}) \right]
\nonumber \\
&& \left.  - 2 c_V c_A
\left[ (k_f \cdot p_f)(k_i \cdot b_{\alpha})  + (k_f \cdot b_{\alpha} )(k_i
\cdot p_f) \right] \right\} ,
\\
\tldW_f/ m_e &=& \delta_{\alpha e} \left\{
 c_V^2 \left[
\left( k_f \cdot ( p_i - p_f) \right) (k_i \cdot b_{\beta})
- \left( k_i \cdot (p_i - p_f ) \right)
( k_f \cdot b_{\beta} ) \right] \right.
\nonumber \\
&& + c_A^2 \left[
\left[k_f \cdot ( p_i + p_f) \right] (k_i \cdot b_{\beta})
- \left[k_i \cdot (p_i + p_f) \right] (k_f \cdot b_{\beta}) \right]
\nonumber \\
&& \left. - 2 c_V c_A \left[ (k_i \cdot p_i) (k_f \cdot b_{\beta} )
+ (k_f \cdot p_i )(k_i \cdot b_{\beta} ) \right] \right\} ~~,
\end{eqnarray}
and
\begin{equation}
b_{i,f} = b_e(k_f) = \frac{m_e}{|\vp_{i,f}|} a_e (p_{i,f}) .
\end{equation}
In the actual calculation we take the limit of $m_e \rightarrow 0$,
keeping $\mu_e {\tilde W}_{i.f}$ and $b_{i,f}$ finite.

The second term in Eq.~(\ref{Dsig}), $\Delta \sigma_{el}$, is calculated as
\begin{eqnarray}
&&\left[ \frac{ G_F^2 B}{16 \pi^5 |\vq| |\vk_i||\vk_f|} \right]^{-1}
 \frac{~d^3}{d k^3_f} \Delta \sigma_{el}
\nonumber \\
& \approx &
\sum_{\alpha,\beta} n^\prime_{l}(|\vk_f|)
\int \frac{d^3 p_i}{|\vp_i| E^*_{\alpha} E^*_{\beta}} \delta(t-t_p)
(E^*_{\alpha} + \omega)  n_{\alpha} (E_{\alpha}^* + U_0(\alpha))
\left[ 1- n_{\beta}(E_{\beta}^* + U_0(\beta)) \right] \tldW_{e}
\nonumber \\ & + & \sum_{\alpha,\beta}
 \left[ 1 - n_{l} (|\vk_f|)  \right]
 \int \frac{d^3 p_i}{ |\vp_i| E^*_{\alpha} }\delta(t-t_p)
 n_{\alpha}(E_{\alpha}^* + U_0(\alpha)) n_{\beta}^\prime(E_{\beta}^{*} + U_0(\beta) ) {\tldW}_{e}
\nonumber \\ & - & \sum_{\alpha,\beta}
 \left[ 1 - n_{l} (|\vk_f|)  \right]
 \int \frac{d^3 p_i}{\vp_i^2 E^*_{\alpha} }\delta(t-t_p) (E^*_{\alpha} + \omega)
n_{\alpha}(E_{\alpha}^* + U_0(\alpha) )
\nonumber \\ && \quad\quad\quad\quad\quad\quad\quad\quad \quad\quad
\times  \left[ 1- n_{\beta}(E_{\beta}^* + U_0(\beta) ) \right]
\frac{\partial {\tldW}_{e}}{\partial t}
\label{DsigEL}
\end{eqnarray}
with
\begin{eqnarray}
 \tldW_e &=& \frac{m_e \mu_e}{|\vk_f|} W_e
\nonumber \\
&=& - c_V^2 \left[ (k_i \cdot p_f) ( p_i \cdot b_e )
+ (k_i \cdot p_i) ( p_f \cdot b_e ) - M_{\beta} M_{\alpha} ( k_i \cdot b_e )  \right]
\nonumber \\ &&
- c_A^2 \left[ (k_i \cdot p_f) (p_i \cdot b_e )
+ (k_i \cdot p_i )( p_f \cdot b_e ) + M_{\beta} M_{\alpha} ( k_i \cdot b_e ) \right]
\nonumber \\ &&
+ 2 c_V c_A \left[ (k_i \cdot p_f )( p_i \cdot b_e )
- (k_i \cdot p_i)( p_f \cdot b_e ) \right]~~ ,
\label{Wel}
\end{eqnarray}
where $b_e=m_e a_e(k_f) / |\vk_f|$.

\subsection{Non-Relativistic Limit}

In order to clarify  relativistic effects we take the non-relativistic
 limit, $p_{\alpha} = (M_{\alpha};0),  p_f = (M_{\beta};0), a_{\alpha} = (0,0,0,1)$.
Then the cross-sections become
\begin{eqnarray}
\frac{d^2 \sigma_0}{d k_f d \Omega_f}
&=&  \frac{G_F^2}{16 \pi^5}
\left[ 1 - n_l (|\vk_f|)  \right] \vk_f^2
\left[ (c_V^2 + 3 c_A^2) +
(c_V^2 - c_A^2) \frac{{\vk_i} \cdot {\vk_f} }{|\vk_i| |\vk_f|} \right] R_1~~,
\label{CrNR0}
\\
\frac{d^2 \Delta \sigma_M}{d k_f d \Omega_f}
&=&   \frac{G_F^2}{16 \pi^5} B
\left[ 1 - n_l (|\vk_f|)  \right] \vk_f^2
\left\{ \cos \theta_{i} \sum_{\alpha,\beta} \left[
\mu_{\alpha} c_A (c_V + c_A)R_2
- 2 \mu_{\beta} c_A(c_V - c_A) R_3  \right] \right.
\nonumber \\
& & \left. + \sum_{\alpha,\beta} \cos \theta_f \left[
\mu_{\alpha} c_A (c_V - c_A) R_2 + 2\mu_{\beta} c_A ( c_V + c_A) R_3 \right] \right\}~~
\label{CrNRM}
\end{eqnarray}
with
\begin{eqnarray}
R_1 &=& \int d^3 p \delta (|\vk_i|- |\vk_f| + E_\alpha(\vp)  - E_\beta(\vp+\vq))
n_\alpha(E_\alpha)[1-n_{\beta}(E_\beta))] ~,
\\
R_2 &=&  \int d^3 p
\delta (|\vk_i|- |\vk_f| + E_\alpha(\vp)  - E_\beta(\vp+\vq))
\nonumber \\ && \quad\quad\quad \times
\left\{ n_\alpha^\prime (E_\alpha) [1 - n_{\beta}(E_\beta)]
+ n_\alpha (E_\alpha) n_{\beta}^\prime(E_\beta) \right\} ~,
\\
R_3 &=&  \int d^3 p
\delta (|\vk_i|- |\vk_f| + E_\alpha(\vp)  - E_\beta(\vp+\vq))
 n_\alpha(E_\alpha) n^\prime_{\beta} (E_\beta) ~,
\end{eqnarray}
where $E_\alpha$ is the single particle energy of particle $a$,
and $\theta_i$ and $\theta_p$ are the polar angles of the initial and
final leptons.

Lai and Qian \cite{lai98} made a further approximation with the long
wave length limit $|\vk_i|-|\vk_f| \rightarrow 0$, and
made $R_1$, $R_2$ and $R_3$ independent of $\theta_i$ and $\theta_f$.
Then, $\sigma_M$ is a linear function of $\theta_i$ and
$\theta_f$.  
This makes is possible to solve the Boltzmann equation  analytically.
However,  this approach does not include the effects of Fermi motion and
cannot be used for the electron contribution
because its mass is taken to be zero. 
Therefore, this approximation is only valid in the very low density regime,
$\rho_B \lesssim 0.1 \rho_0$.

\newpage

\section{Results and Discussion of Neutrino Cross-Sections}

In this section we present the cross-sections for neutrino scattering
($\nu_e \rightarrow \nu_e$) and absorption ($\nu_e \rightarrow e^{-}$)
in matter with and without a magnetic field.
We set the lepton fraction to be fixed as $Y_L = 0.4$, and  the neutrino incident
energy is taken to be its chemical potential, $|\vk_i| = \epsi_\nu$,
unless otherwise noted.
In Eq.~(\ref{elLag}) we utilize the parameters for the weak-interaction, 
$c_V$ and $c_A$ from Ref.~\cite{rml98}.

\subsection{Neutrino Cross-Sections without a Magnetic Field}

\begin{figure}[ht]
{\includegraphics[scale=0.5,angle=270]{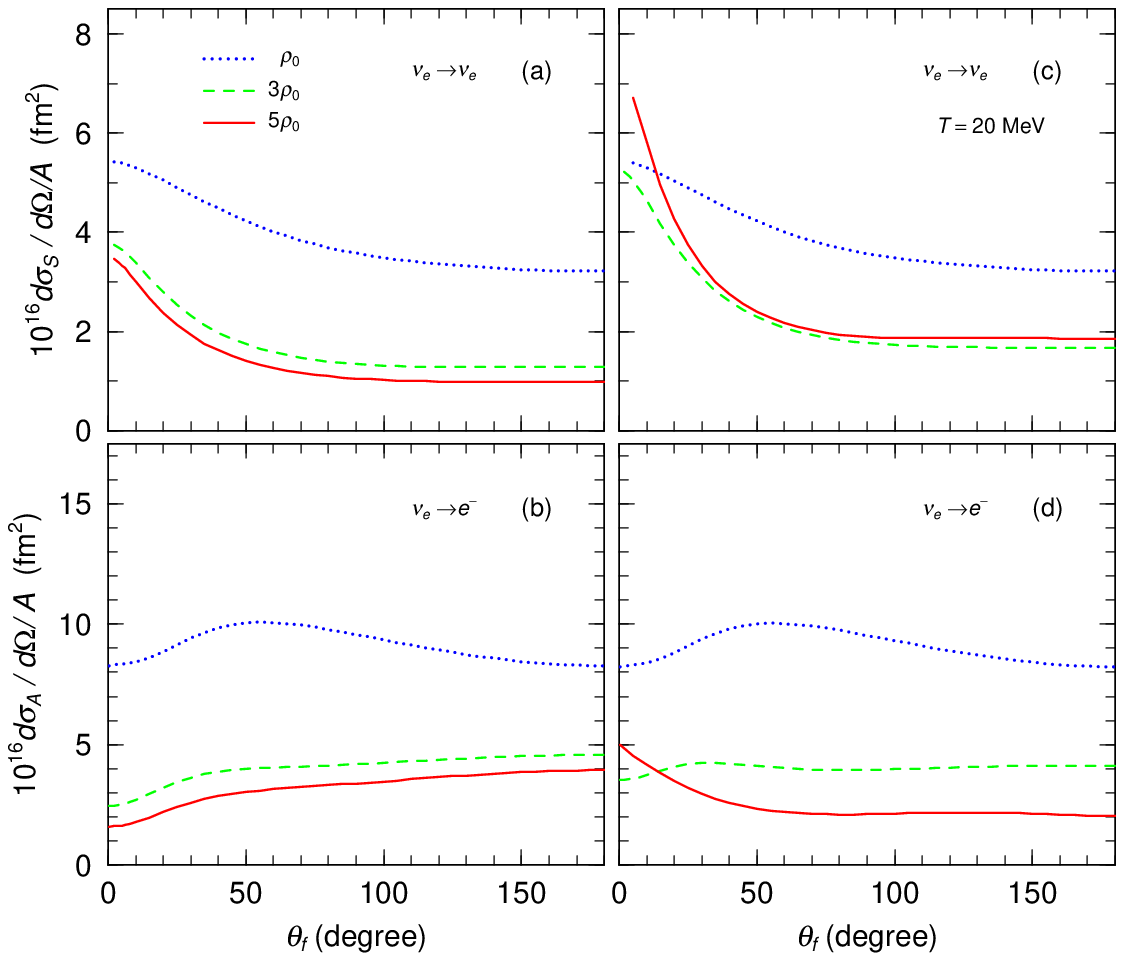}}
\caption{\small (Color online) Density dependence of the scattering (a and c) and
absorption differential cross-sections (b and d) of  neutrinos in
neutron-star matter at $T=20$~MeV without $\Lambda$s (a and b)
and with $\Lambda$s (c and d).
The initial neutrino angle is taken to be $\theta_i=0^\circ$.
Dotted, dashed and solid lines represent the results for
$\rho_B =$ $\rho_0$, $3\rho_0$ and $5 \rho_0$, respectively.}
\label{dfCrT20}
\end{figure}

In Fig.~\ref{dfCrT20} we show the density dependence of the
differential cross-section per baryon at $T=20$ MeV for the scattering  (a and c)
and absorption (b and d) of  neutrinos in matter without $\Lambda$s
(a and b) and with $\Lambda$s (c and d).
The subscripts  'S' or 'A' refer to the scattering or
absorption cross-sections, respectively.
Solid and dashed lines show the results in  matter
including $\Lambda$s or no $\Lambda$s, respectively.
We see that the scattering cross-sections
are forward peaked, while the absorption cross-sections decrease at forward angles when $\rho_B \le 3\rho_0$.

\begin{figure}[ht]
{\includegraphics[scale=0.5,angle=270]{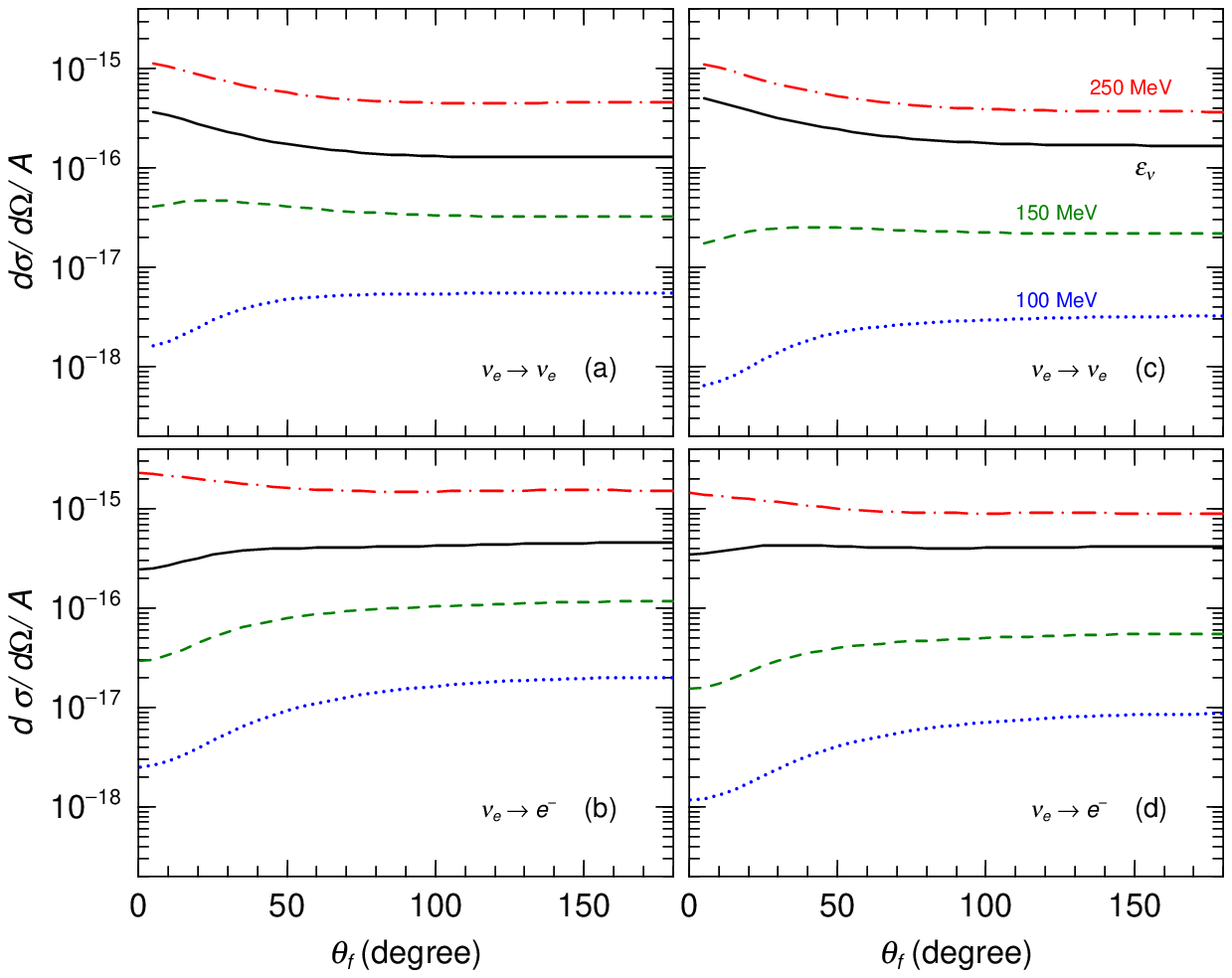}}
\caption{\small (Color online)
Scattering (a and c) and absorption differential cross-sections
(b and d) for neutrinos for $\theta_i=0$
versus the final lepton angle $\theta_{f}$
with various incident neutrino energies in PNS
matter at $\rho_B = 3 \rho_0$ and $T = 20$ MeV.
Dotted, dashed and dot-dashed lines show the results
for incident neutrino energies $|\vk_i| =$ 100, 150, and  250 MeV, respectively.
Solid lines show results for the incident neutrino energy
equal to the neutrino chemical potential, i.e.~$|\vk_i| =\epsi_{\nu}$.}
\label{CrEnT20}
\end{figure}

In Fig.~\ref{CrEnT20}, we show the energy dependence of the differential
cross-sections per baryon  at $\rho_B = 3\rho_0$ and $T=20$ MeV
for various incident neutrino energies.
The solid lines show  the results for the incident neutrino
energies equal
to the neutrino chemical potentials, i.e.~$|\vk_i| =\epsi_{\nu} $.
Dotted, dashed and dot-dashed lines represent the results at
$|\vk_i|=$100, 150 and 250 MeV, respectively.

For $|\vk_i| = 100$ MeV,
the cross-sections show a minimum at forward angles.
With the increase of incident energy, however, the cross-sections gradually
become larger and finally become peaked at forward angles.
This behavior arises from the the difference in Fermi distributions
between the spin-up and spin-down particles,
as was discussed in Ref.~\cite{MKYCR11}.
This Pauli blocking affects the results at all angles, and,
in particular,  manifests itself at forward angles.
However, this Pauli blocking effect becomes smaller at higher incident energies
as shown in Fig.~\ref{CrEnT20}.
We have confirmed that the cross-sections always show forward peaks
when we turn off the Pauli blocking term for the final lepton,
$(1 -n_{l})$.
We can therefore conclude that the Pauli blocking effect is clearly exhibited
at  low incident energy as a  suppression of  the differential
cross-sections at  forward angles.

\subsection{Differential Neutrino Cross-Sections in a  Magnetic Field}

In this subsection we discuss effects of a magnetic field on
the neutrino reactions in neutron-star matter.
For illustration, we first calculate the differential cross-sections per baryon,
$d \sigma_{S,A} / d \Omega / A$ with an initial neutrino angle of
$\theta_i=0^\circ$ at a matter density of $\rho_B = 3 \rho_0$
and a  magnetic field of $B=2 \times 10^{17}$G.
This gives $\mu_N B = 0.63$ MeV, where $\mu_N$ is the nuclear magneton.
The initial momenta are taken to be equal to
the chemical potential in each case, {\it i.e.} $|\vk_i| = \epsi_{\nu}$.

\begin{figure}[ht]
{\includegraphics[scale=0.55,angle=270]{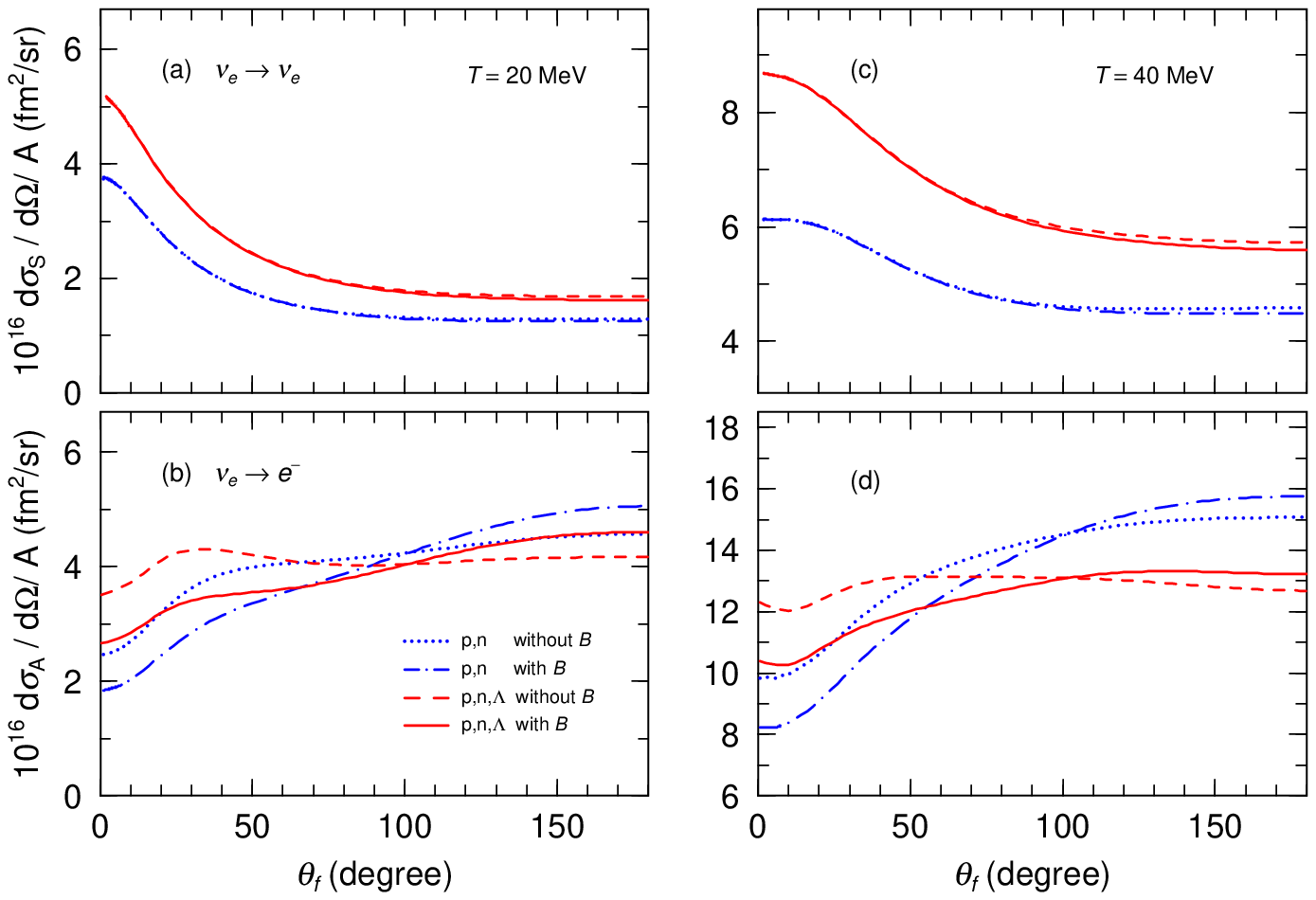}}
\caption{\small (Color online) Effects of magnetic fields on the  differential cross
sections per baryon $d \sigma/ d \Omega/$A, from
Eq.~(\ref{CrsNM}) in units of $10^{-16}$ fm$^2$.
This figure is  obtained at a density of $\rho_B = 3 \rho_0$
at $T$ = 20 MeV (a and b) and 40 MeV (c and d).
Upper (lower) panels are for neutrino scattering (absorption).
Initial momentum and angle of incident neutrinos are taken
to be $|\vk_i| = \epsi_{\nu}$ and $\theta_i=0^\circ$.
Solid and short-dashed lines represent results
including $\Lambda$s with and without
a magnetic field $B=2 \times 10^{17}$G, respectively.
Dot-dashed and dotted lines represent results without $\Lambda$s. }
\label{CrAngT}
\end{figure}

In Fig.~\ref{CrAngT} we show the neutrino scattering
($\nu_e \rightarrow \nu_e$) cross-sections in the upper
panels (a and c) and the absorption ($\nu_e \rightarrow e^{-}$)
cross-sections in lower panels (b and d), in Eq.~(\ref{CrsNM}).
Left and right panels are for temperatures
$T=20$ and 40 MeV, respectively.
Solid and dashed lines show the results with and w/o $\Lambda$s, respectively.
For reference, we also plot results without a magnetic field for both cases, and including
 (dot-dashed lines) and excluding (dotted lines) $\Lambda$s.

\begin{figure}[ht]
{\includegraphics[scale=0.55,angle=270]{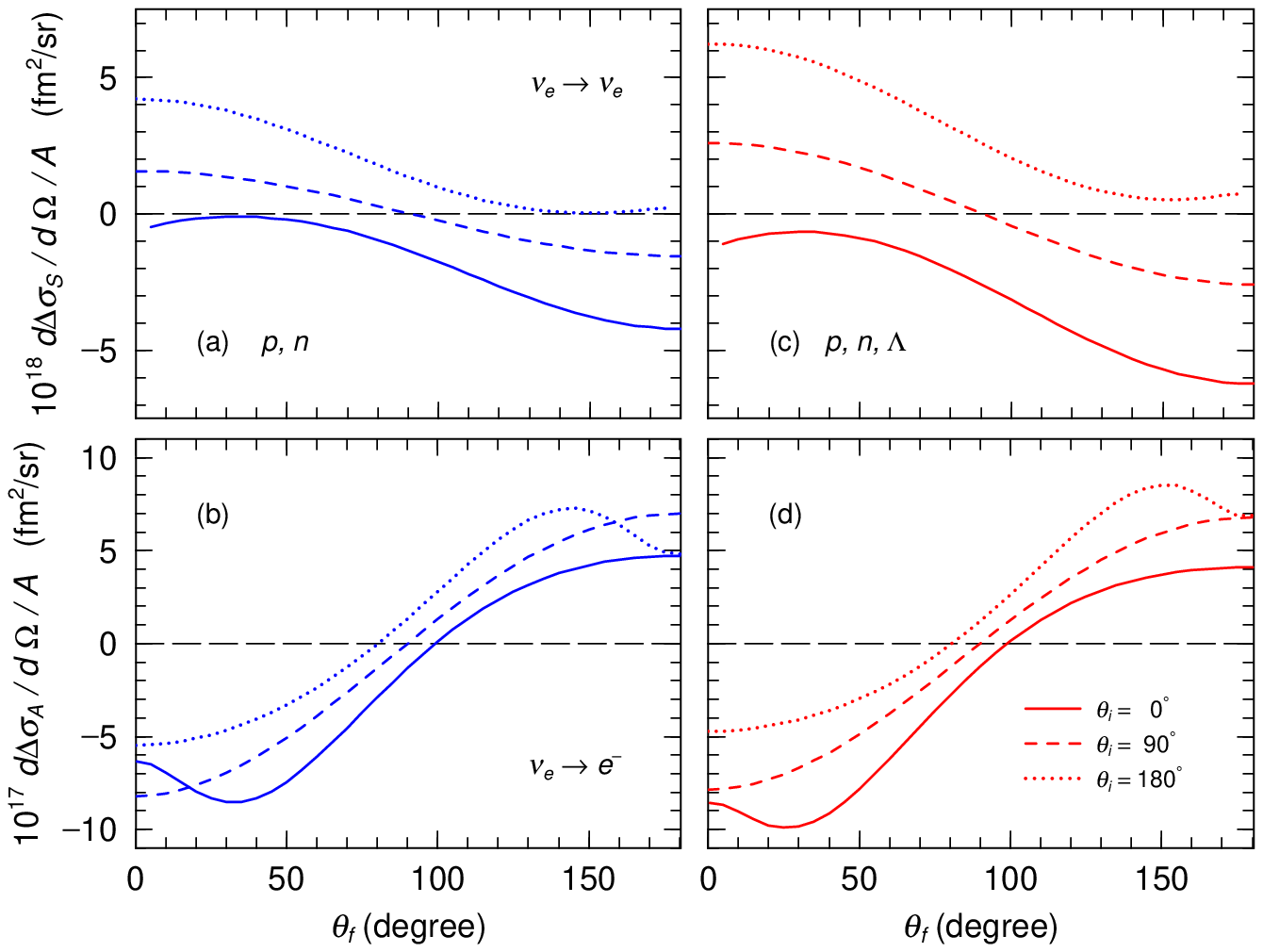}}
\caption{\small (Color online) Magnetic parts of the differential cross-sections per
baryon, $d \Delta \sigma / d \Omega/$A, corresponding to the
2nd term in Eq.~(\ref{CrsNM}), in units of $10^{-16}$ fm$^2$ for
neutrino scattering ($\nu_e \rightarrow \nu_e$) at $\rho_B = 3 \rho_0$ for $T$ = 20 MeV without $\Lambda$s (a) and with $\Lambda$s (c). Lower panels (b and d) are
the same as the upper panels but for  neutrino absorption
($\nu_e \rightarrow e^{-}$) in a system without (b) and
 with  $\Lambda$s (d). Initial momentum and angle of the incident
neutrinos are taken to be $|\vk_i| = \epsi_{\nu}$. Solid, dashed
and dotted lines represent results for
$\theta_i=0^\circ$, $90^\circ$ and $180^\circ$, respectively. }
\label{MgCrAgT}
\end{figure}

This figure beautifully indicates that the magnetic field does not much affect 
 the scattering cross-sections when $B \approx 2 \times 10^{17}$G.
Actually the contributions from each individual particle such as
protons and neutrons are not so small.
These contributions, however,  tend  to cancel each other out.
However, the magnetic field suppresses the absorption cross-section
in the forward direction and enhances it in the backward direction.
In particular, near $\theta_f \approx 0^\circ$, the suppression
from the magnetic field is as much as  20 $-$30 \%.
This contribution is almost as large as that from the $\Lambda$ particles.

In Fig.~\ref{MgCrAgT} we show the magnetic parts of the differential
cross-sections, $\Delta \sigma$ of Eq.~(\ref{CrsNM}), at $\rho_B=3\rho_0$ and $T=20$ MeV. Upper panels  are for
neutrino scattering ($\nu_e \rightarrow \nu_e$)  and lower panels (b and d) are for absorption ($\nu_e \rightarrow e^{-}$).
Right and left panels are for
matter including and excluding $\Lambda$s, respectively.
Solid, dashed and dotted lines represent results
for  incident angles, of  $\theta_i=0^\circ$,
$90^\circ$ and $180^\circ$, respectively.
In these calculations we keep the difference of the azimuthal angle
between the initial and final leptons equal to  zero, i.e.~$\phi_f - \phi_i=0$.

These calculations show that the magnetic field enhances
the scattering cross-sections in the direction along the magnetic field
(arctic direction).
For absorption an enhancement appears in the opposite direction (antarctic direction).
These asymmetries of
the scattering and absorption cross-sections of neutrinos by the magnetic field would lead to the  coherent effect of enhancing
the neutrino drift in the arctic direction while  suppressing it in the
antarctic direction, as will be discussed below.

\subsection{Angular-integrated Neutrino Cross-Sections in a  Magnetic Field and The Asymmetries}

In order to discuss the effects of neutrino transfer inside the PNS at subsection E,
we here calculate the scattering cross-sections  integrated over the
momenta of the initial neutrinos
\begin{eqnarray}
\sigma_{S} (|\vk_f|, \theta_f) &=&
\sigma_{S}^0 (|\vk_f|, \theta_f) + \Delta \sigma_{S} (|\vk_f|, \theta_f)
\nonumber \\ &=& \frac{1}{\rho_B}
 \int \frac{d^3 k_i }{(2 \pi)^3}
n_{\nu}(|\vk_i|)
\frac{d^3 \sigma_S^0 (\vk_i, \vk_f)}{d k_f^3~~~~}
+ \frac{1}{\rho_B} \int  \frac{d^3 k_i }{(2 \pi)^3} n_{\nu}(|\vk_i|)
\frac{d^3 \Delta \sigma_S (\vk_i, \vk_f)}{d k_f^3~~~~} ~. ~~
\label{DsigS}
\end{eqnarray}
The absorption cross-sections are however integrated
 over the momenta of the final electrons as
\begin{eqnarray}
\sigma_{A} (|\vk_i|, \theta_i) &=&
\sigma_{A}^0 (|\vk_i|, \theta_i) + \Delta \sigma_{A} (|\vk_i|, \theta_i)
\nonumber \\ &=&
 \int d^3 k_f \frac{d^3 \sigma_A^0 (\vk_i, \vk_f)}{d k_f^3~~~~}
+  \int d^3 k_f
\frac{d^3 \Delta \sigma_A (\vk_i, \vk_f)}{d k_f^3~~~~} ~~.
\end{eqnarray}
Note that  the non-magnetic parts of the integrated cross-sections,
$\sigma^0_{S,A}$, are also integrated the same way.

\begin{figure} 
\hspace*{0.5cm}
{\includegraphics[scale=0.5,angle=270]{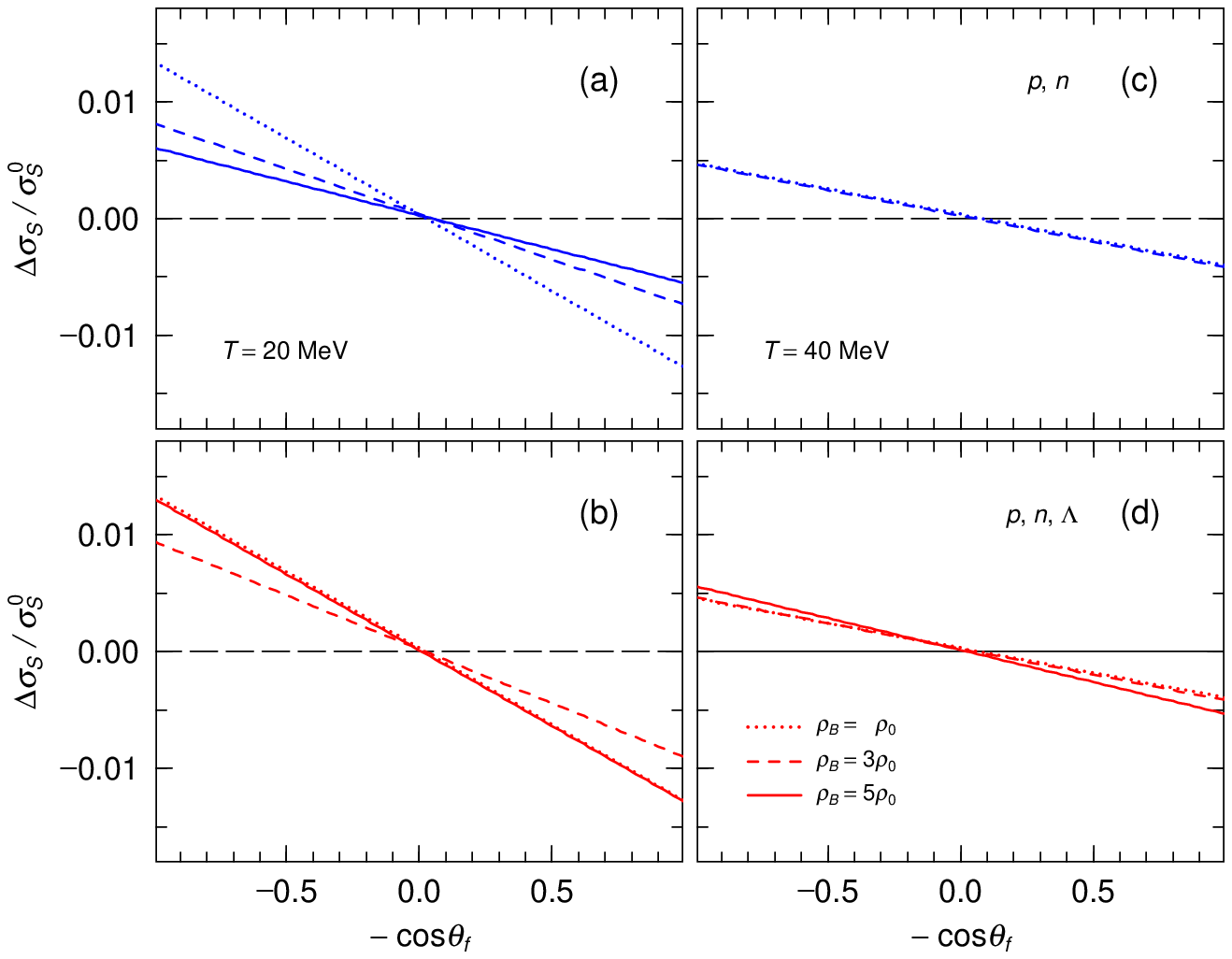}}
\caption{\small (Color online) Ratios of the magnetic part of the scattering
cross-sections ($\Delta \sigma_S$) to the cross-sections without a
magnetic-field ($\sigma_S^0$) without $\Lambda$s (a and c) and
 with $\Lambda$s (b and d) at $T=20$ MeV (a and b) and at $T=40$ MeV
(c and d).
Dotted, dashed and solid lines represent the results for densities of
$\rho_B = \rho_0$, $3 \rho_0$ and $5 \rho_0$, respectively.}
\label{TCrScMg}
\end{figure}

\begin{figure}
{\includegraphics[scale=0.5,angle=270]{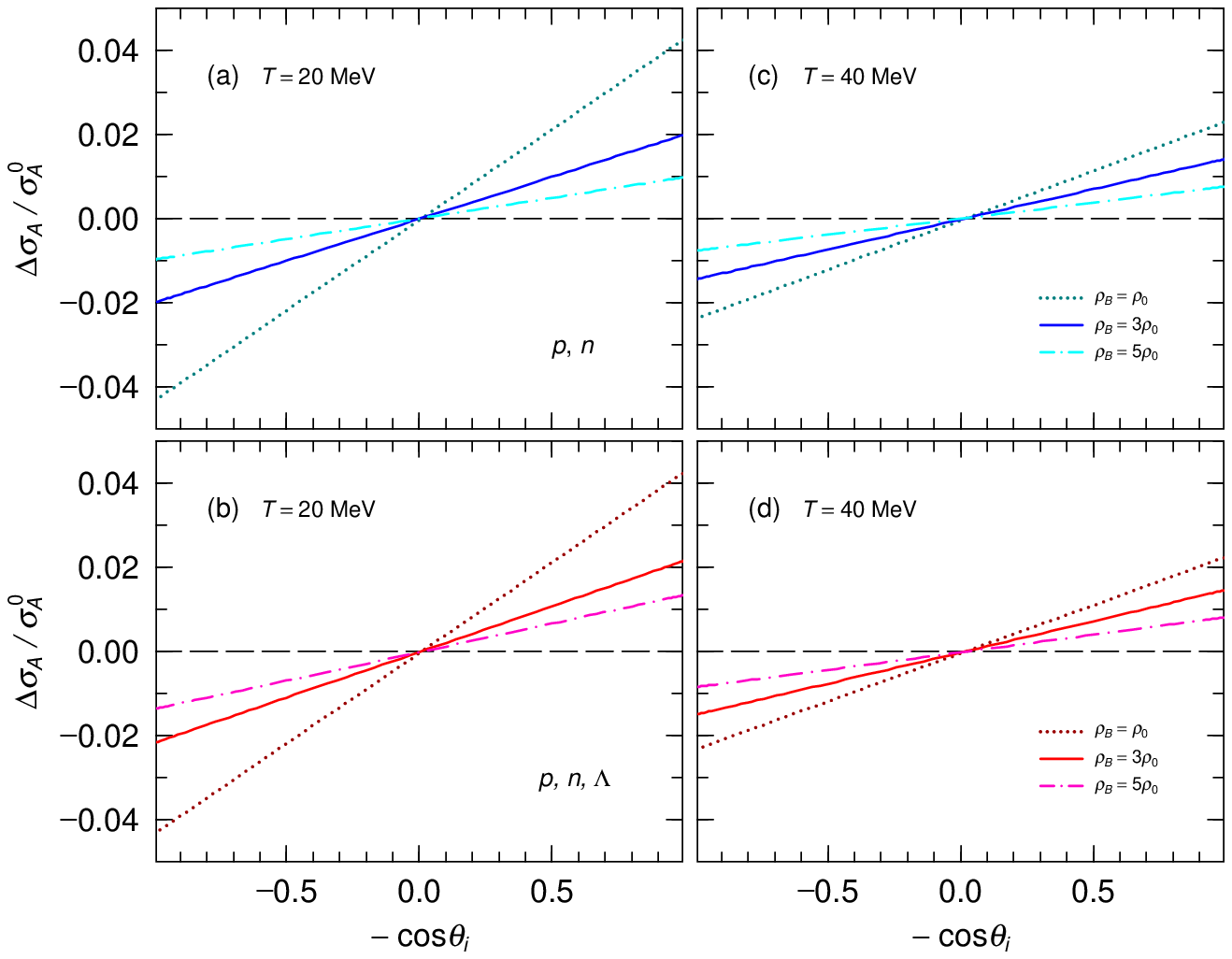}}
\caption{\small (Color online) Ratios of the magnetic part of the absorption
cross-sections ($\Delta \sigma_A$) to the cross-sections
without a magnetic-field ($\sigma_A^0$) without $\Lambda$s
(a and c) and on matter with $\Lambda$s (b and d) at $T=20$ MeV (a and b)
and at $T=40$ MeV (c and d).
Dotted, dashed and solid lines represent the
results for $\rho_B = \rho_0$, $3 \rho_0$ and $5 \rho_0$,
respectively. }
\label{TCrAbMg}
\end{figure}

Figures \ref{TCrScMg} and ~\ref{TCrAbMg} show $\Delta \sigma_S /
\sigma_S^0$ with $|\vk_i| = \epsi_{\nu}$ and $\Delta \sigma_A
/\sigma_A^0$ with $|\vk_f| = \epsi_{\nu}$ as functions of
$\theta_f$ and $\theta_i$, respectively, for matter
densities, $\rho_0 \le \rho_B \le 5 \rho_0$.
We plot results for matter without $\Lambda$s (upper panels)
and with $\Lambda$s (lower panels) at $T = 20$ MeV (left panels) and  $T = 40$
MeV (right panels).
Similar to the differential cross-sections,
the magnetic field enhances the integrated scattering cross-sections
and suppresses the integrated absorption cross-sections
in the arctic direction parallel to the magnetic field ${\bm B}$.
The magnetic field has an opposite effect in the anti-parallel antarctic
direction.  Therefore, we may conclude that a magnetic field increases the neutrino emission
in the arctic direction and decreases it in the antarctic direction.

\begin{figure}[ht]
\begin{center}
\includegraphics[scale=0.5,angle=270]{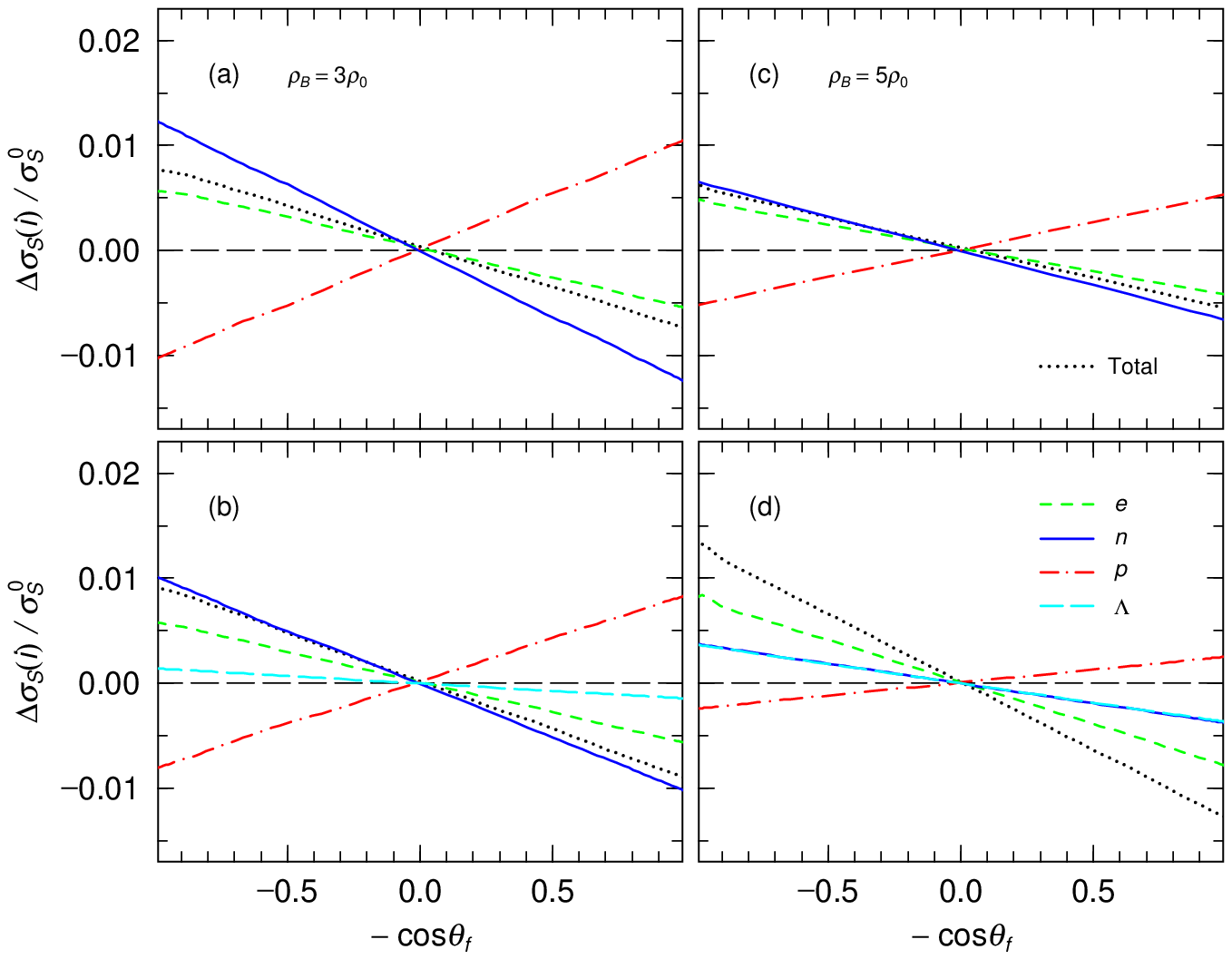}
\caption{\small (Color online) Contributions from each constituent particle to the
magnetic part of the scattering cross-sections $(\Delta {\sigma}_S)$ at $T = 20$ MeV
divided by the integrated cross-sections without a magnetic-field ($\sigma^0_S$). Upper(lower) panels exhibit the results without (with) $\Lambda$s
at $\rho_B = 3 \rho_0$ (left) and  $\rho_B = 5 \rho_0$ (right), respectively. Dashed, solid, dot-dashed and long dashed
lines represent contributions from electrons, neutrons, protons and
$\Lambda$s, respectively.
Dotted lines represent a sum of the contributions.
In panel (d), solid and long dashed lines are indiscernible.
} \label{TScPT1}
\end{center}
\end{figure}

In Fig.~\ref{TScPT1}, we show the contribution of each constituent particle to the
scattering cross-sections without $\Lambda$s (upper panels) and with $\Lambda$s (lower panels)
at $\rho_B = 3\rho_0$
(left panels) and $\rho_B = 5 \rho_0$ (right panels).
Only the contribution from the protons is
opposite to those from electrons, neutrons and $\Lambda$s because of the
different signs of the magnetic moments.
These contributions tend to cancel to each other, and
the magnetic parts of the scattering cross-sections become slightly smaller.
However, when one allows $\Lambda$s to appear in the system,
the proton fraction decreases and in this case the cancellation is not as
large as the case without $\Lambda$s (see Fig.~\ref{EOSf}).

\begin{figure}[ht]
\hspace*{0.5cm}
{\includegraphics[scale=0.5,angle=270]{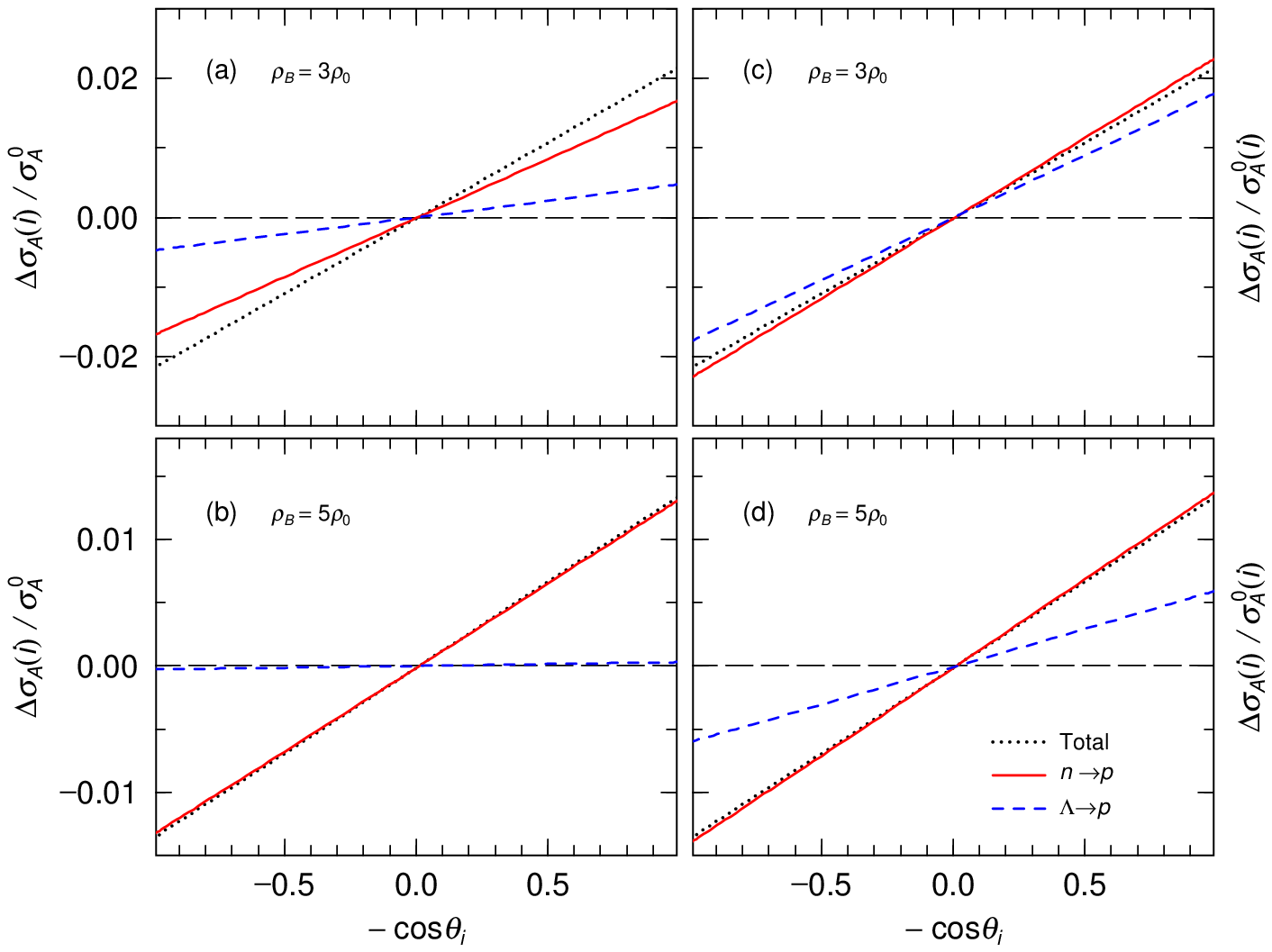}}
\caption{\small (Color online) Contributions from each constituent particle in the
magnetic part of the absorption cross-sections $(\Delta \sigma_A)$ at $T = 20$ MeV with
$\Lambda$s  at $\rho_B =3 \rho_0$ (upper) and $\rho_B =5 \rho_0$ (lower panels). Left  and right panels
exhibit results divided by the non-magnetic parts of the total
cross-sections $(\sigma_A^0)$ and  their non-divided respective
contributions  $(\sigma_A^0 (i))$. Solid and dashed lines represent the contributions
from the $n \rightarrow p$ and $\Lambda \rightarrow p$ processes,
respectively. } \label{TAbPT}
\end{figure}

Solid and dashed lines in Fig.~\ref{TAbPT}  show the contributions from the $n \rightarrow
p$ and $\Lambda \rightarrow p$  neutrino absorption processes,
 respectively. Upper and lower panels exhibit the
results at $\rho_B = 3\rho_0$ and $\rho_B = 5 \rho_0$,
respectively. Results in the left and right panels are
divided by the non-magnetic parts of the integrated cross-sections and their
respective non-magnetic contributions.

Contributions from the $\Lambda \rightarrow p$ process seem to be much
smaller than those from the $n \rightarrow p$ in the left panels,
but in the right panels the former contributions are
as large as the latter.
This  apparent difference is because of the small Cabibbo angle,
$\sin^2 \theta_C \approx 5.0 \times 10^{-2}$.
Since the non-magnetic part of the $\Lambda \rightarrow p$ process
is associated with a strangeness change of $\Delta S =1$,
its transition probability is $\sim \sin^2 \theta_c$ times smaller than
that of the $n \rightarrow p$, $\Delta S = 0$, process.
As a result, contributions from the $\Lambda \rightarrow p$ process to
the total non-magnetic part becomes very small.
However, when one divides the small contributions by the small quantities from respective non-magnetic
parts, the ratio shows an interesting difference as illustrated  in the right panels.
With $\Lambda$s present, the proton fraction becomes
smaller as the density changes, and the contribution from the magnetic parts
of the $\Lambda \rightarrow p$ process becomes remarkably larger.

\subsection{Neutrino Mean-Free-Paths}

\begin{figure}[ht]
\begin{center}
{\includegraphics[scale=0.55,angle=270]{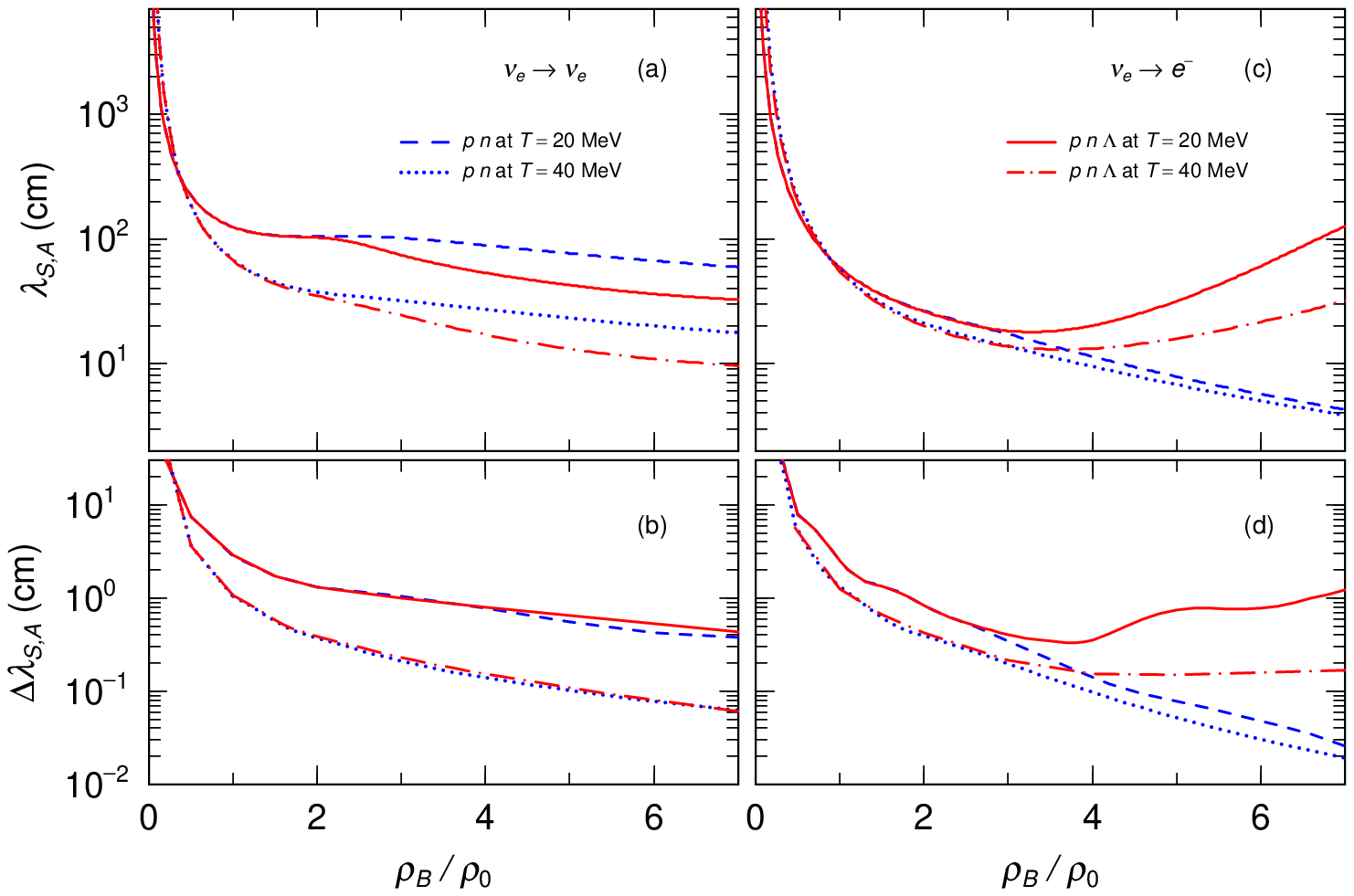}}
\caption{\small
(Color online) Upper panels show the neutrino MFP for
 scattering (a) and absorption (c) without a magnetic field.
The lower panels show the magnetic contribution to
the MFP for scattering (b) and absorption (d). Here the neutrino incident energy is fixed as the chemical potential.
Since the magnetic part of the MFP for  scattering is negative,
we multiply by $(-1)$.
Solid and dashed lines represent the results at $T=20$ MeV with and without $\Lambda$s, respectively.
Dot-dashed and dotted lines represent the results at $T=40$ MeV with and without $\Lambda$s, respectively.  }
\label{MFpath}
\end{center}
\end{figure}

In order to apply the above results to
astronomical phenomena, we discuss the neutrino mean-free-paths (MFPs).
In Fig.~\ref{MFpath}, we show the density dependence of the
neutrino MFPs, $\lambda_{S,A} = V/\sigma_{S,A}$ with the system volume V, for the scattering (a)
and the absorption (c) processes at $T= 20$ and 40 MeV for $B=0$.
For this illustration, the incident neutrino energy is fixed to be equal to its chemical potential.

The scattering and absorption MFPs rapidly decrease as
the density increases  up to $\rho_B \approx (2-3) \rho_0$.
When the system does not include $\Lambda$s, both MFPs (dashed and dotted lines)
decrease monotonically even beyond $\rho_B \approx (2-3) \rho_0$.
When the system includes $\Lambda$s, the scattering
MFPs also decrease, but the absorption MFPs increase in  $\rho_B \gtrsim 3 \rho_0$, because the cross-sections
for  $\nu_e +  \Lambda \rightarrow p + e^{-}$ are  smaller than those
of $\nu_e + n \rightarrow p + e^{-}$.

In addition, we show the magnetic contributions to the MFPs,
$\Delta \lambda_{S,A} \equiv [V/\sigma(0^\circ) - V/\sigma(180^\circ)]/2$,
 in the lower panels (b and d).
We should note that the $\sigma$ contribution from the scattering
process is calculated by an integration over final angle, which is
not the same as $\sigma_S$ defined in Eq.~(\ref{DsigS}). We see, again, that  the contribution of the magnetic field is $\sim 1-2$\% of
the non-magnetic parts.

The slopes of the magnetic parts of the neutrino cross-sections
$\Delta \sigma_{S,A}$ are almost constant as a function of
$\cos \theta_{i,f}$ (see Figs.~\ref{TCrScMg} and \ref{TCrAbMg}).
Hence
if we define the slopes as $S_{S,A} = ( \Delta \sigma_{S,A} /
\sigma_{S,A}^0 ) / \cos \theta_{i,f} $, the integrated cross-sections
$\Delta \sigma_{S,A}$ can be  approximately written as
\begin{equation}
\sigma_{S,A} \approx \sigma_{S,A}^0 (1 + S_{S,A} \cos \theta_{i,f} ) .
\label{SigAng}
\end{equation}
The discrepancy in the use of this formula is estimated to be less than 1 \%.

\begin{figure}[ht]
\begin{center}
\vspace{-1.5em}
{\includegraphics[scale=0.45]{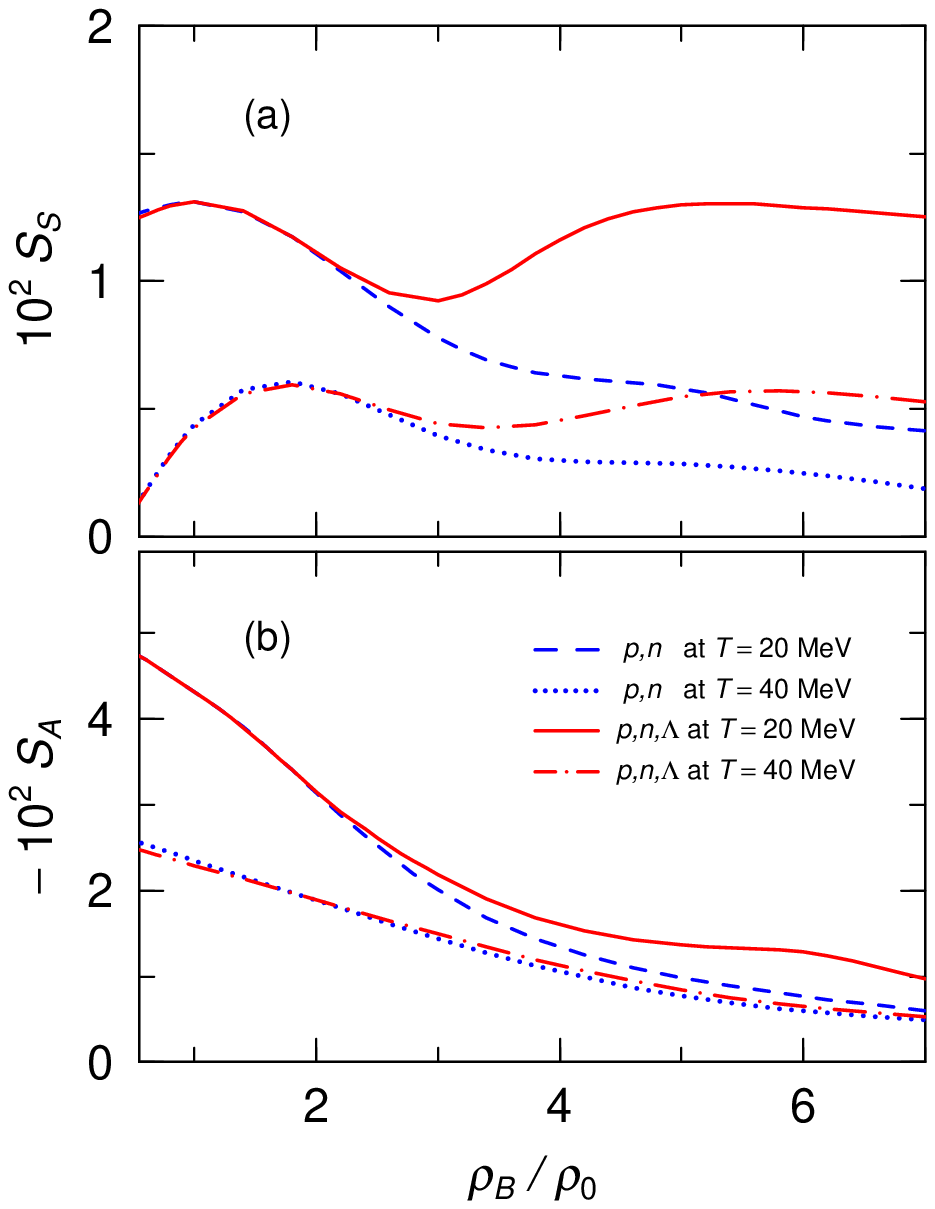}}
\caption{\small (Color online) Density dependence of $S_S$ when $|\vk_{i}| =
\epsi_\nu$ (a) and $S_A$ when $|\vk_{f}| = \epsi_\nu$ (b). 
Various lines show the results without $\Lambda$s  at $T = 20$ MeV Dashed)  and without $\Lambda$s at $T=40$ MeV(dotted),
with $\Lambda$s at $T=20$ MeV (solid)  and with $\Lambda$s at $T=40$ MeV (dot-dashed). }
\label{ScPar}
\end{center}
\end{figure}

Since $S_{S} > 0$ and  $S_{A} < 0$, the neutrinos scatter and absorb
in the arctic direction due to the magnetic field.
In Fig.~\ref{ScPar}, we show the density dependence of $S_S$ (a) and
$S_A$ (b). It is evident that the effects of the magnetic field become
smaller as the temperature and the density increase.
This density dependence arises from the fact that $\Delta \sigma$ is
approximately proportional to the fractional area of the distorted
Fermi surface caused by the magnetic field.  Hence, the relative
strength $\Delta \sigma_{S,A}/\sigma^0_{S,A}$ diminishes  with  increasing density.

However, the density dependence of $S_S$ including
$\Lambda$s exhibits a local minimum  around $\rho_B \approx 3 \rho_o$
and increases again
in the density region, $3\rho_0 \lesssim \rho_B \lesssim (5 - 6) \rho_0$.
As commented before, the Lambda fraction rapidly increases for
$\rho_B \gtrsim \rho_0$, and its contribution enhances $S_S$
(see Fig.~\ref{EOSf} and Fig.~\ref{TScPT1}).


\subsection{Neutrino Transport and Pulsar Kick Velocities}

Next we discuss implications of these findings
for neutrino transport in a strongly magnetized PNS.
It has been pointed out that  asymmetric neutrino emissions
may cause the pulsar kicks of magnetars~\cite{pac92,mag3}.
Most of the explosion energy is emitted as neutrinos.
In this subsection, we estimate the momentum transfer
from the asymmetric neutrino emission.

In the interpretation of actual phenomena, many different effects may
contribute to the generation of pulsar kicks.  
One must, therefore, solve the time evolution of PNSs
with a numerical simulation.
However, our purpose is to examine qualitatively
the effects from our asymmetric cross-sections on the kick velocity. 
Therefore, we we can limit our discussion to only  effects of the
asymmetric cross-section discussed above.

For this purpose we can assume that the PNS   is in local equilibrium,
and that the neutrinos propagate through the dense nuclear matter in the
presence of a  strong magnetic field, and that eventually the neutrinos 
are emitted asymmetrically.
Along with these assumptions, we ignore the effects of other neutrino 
processes, such as the direct and moderate URCA processes
\cite{dorofeev,kisslinger2,kisslinger3}, and also the momentum transfer
to the medium at each local position.

The time scale for PNS evolution is  much larger than
that of the emitted neutrino propagating inside the PNS.
Therefore, to estimate the neutrino momentum transport, 
we can conjecture that PNS is static, and that the neutrino transfer
makes a continuous  current in the equilibrium matter. Furthermore, we
simplify  the PNS as having a fixed temperature and magnetic field. 
These simplifying assumptions for the purpose of this work, 
which is to qualitatively examine effects of a magnetic field 
on the PNS momentum. 
Clearly, more investigation beyond the present assumptions is warranted 
and will be the subject of future work as discussed in Sec. VI.

\subsubsection{Boltzmann Equation}

We start with the phase-space neutrino distribution function
$f_\nu(\vbr,\vk)$ and calculate the asymmetric neutrino emission
from the $f_\nu$ function.
This $f_{\nu}$ satisfies the following Boltzmann equation
\begin{equation}
\left( \frac{\partial}{\partial t} +  {\hat k}
\cdot \frac{\partial}{\partial \vbr} \right) f_\nu(\vbr,\vk) = I_{coll}
\end{equation}
with
\begin{eqnarray}
I_{coll}
&=& \sum_{i,j}
\int \frac{d^3 k_l}{(2 \pi)^3}
 \frac{d^3 p_i}{(2 \pi)^3}  \frac{d^3 p_j}{(2 \pi)^3} W_{if}
 \left\{ f_l (\vk_l) f_j(\vp_2) \left[ 1 - f_\nu(\vk) \right]
 \left[ 1 - f_i (\vp_1) \right]
\right. \nonumber \\ && \left. ~~~~~~~~~~~~~~~~~~~~~~~~~~~~~
-  f_\nu (\vk) f_i(\vp_1) \left[ 1 - f_l (\vk_l) \right]
 \left[ 1 - f_j (\vp_2) \right] \right\} ,
\end{eqnarray}
where
$W_{if}$ is the reaction probability.  The index $l$ denotes
leptons, electrons or neutrinos, and the indices $1$ and $2$ label the
target particles, e.g. baryons and electrons.
In the above equations, we omit the contribution from the neutrino
mean-field because its depth is about a few ten eV
($G_F \rho_0 \approx 15$ eV),
and the magnetic contribution is much less.

Here, we introduce several assumptions to obtain  a solution to the
Boltzmann equation.
First, we assume that the system is almost in equilibrium, and that
$f_\nu (\vbr,\vk)$ can be separated into two parts
\begin{equation}
 f_\nu (\vbr,\vk) =  f_0 (\vbr,\vk) +  \Delta f (\vbr,\vk)
= \frac{1}{1 + \exp[(|\vk| - \epsi_\nu(\vbr))/T]} + \Delta f (\vbr,\vk) ~~,
\end{equation}
where the first and the second terms are the local equilibrium
part and the deviation from the equilibrium, respectively, with the neutrino
chemical potential $\epsi_\nu(\vbr)$ at the position $\vbr$.
The phase-space distribution functions of other particles are assumed to
have local thermodynamic equilibrium distributions. 
In addition, we also omit the contribution from
$e^{-} + B \rightarrow B^\prime + \nu_e$.
The collision term can  thus be  written as
\begin{eqnarray}
I_{coll} &\approx& \sum_{ij}
\int \frac{d^3 k_l}{(2 \pi)^3}
 \frac{d^3 p_i}{(2 \pi)^3}  \frac{d^3 p_j}{(2 \pi)^3} \left( W_{S}
 \biggl\{  \Delta f (\vk_l)  \left[( 1 - f_0(\vk) ) f_i ( 1 - f_j )
-  f_0 (\vk) f_i ( 1 - f_j )  \right]  \right.
\nonumber \\ &&  ~~~~~~~~~~~~~~~~~~~~~~~~~~~~~~~~~
-  \Delta f (\vk) \left[ ( 1 - f_0 (\vk_l) ) f_i ( 1 - f_j)
- f_0 (\vk_l) f_i ( 1 - f_j ) \right]  \biggr\}
\nonumber \\ &&  ~~~~~~~~~~~~~~~~~~~~~~~~~~~ \left.
- W_{A}  \Delta f (\vk)  \left[
f_1  \left( 1 - f_e (\vk_l) \right)
(1 - f_2 ) \right]  \right)~~ ,
\label{IcollA}
\end{eqnarray}
where $W_{S}$ and  $W_{A}$ are the scattering and absorption probabilities.

We make the further assumption that only the absorption process
makes a dominant contribution to the neutrino momentum transport.
When the $I_{coll}$ in Eq.~(\ref{IcollA}) is integrated over $\vk$,
the term proportional to $W_S$, which represents the  contribution from
the scattering, becomes zero, {\it i.e.} this part does not change 
the number of emitted neutrinos. 
The scattering process  enhances the asymmetry,
but the magnetic field contribution to the scattering cross-section is
small.
Hence, the approximation of ignoring the scattering process may slightly
underestimate the asymmetry, but does not significantly change  
the estimated effect. 

By ignoring the scattering contributions, we can treat the
neutrino trajectory as the straight line and
simply express the Boltzmann equation for the neutrino transport as
\begin{equation}
 {\hat k} \cdot \frac{\partial}{\partial \vbr} f_\nu(\vbr,\vk)
= {\hat k} \cdot \frac{\partial \epsi_\nu}{\partial \vbr}
\frac{\partial f_0}{\partial \epsi_\nu}
+  {\hat k} \cdot \frac{\partial \Delta f}{\partial \vbr}
= - \frac{ \sigma_{A} (\vbr,\vk)}{V} \Delta f (\vbr,\vk) ,
\label{Boltz1}
\end{equation}
where the absorption cross-section $\sigma_{A}$ is a function of $\vk$
and $\rho_B(\vbr)$. 

In the present approximation the neutrinos are taken to propagate along 
a straight line, which gives us an analytical solution for the above
Boltzmann equation as explained below. 
First,  we define a plane $A_0$ that  is perpendicular to 
the neutrino momentum $\vk$.  
This plane is constructed to intersect the center of the neutron star, 
which we take to be the origin of the coordinate system $\vbr \equiv (0,0,0)$.
Then, we introduce
$x_L$ and $\vR_T$ such that $\vbr =x_L \vk+\vR_T$, where
$x_L$ is the  component of $\vbr$ parallel to $\vk$ and   $\vR_T
\perp \vk$.  In terms of $x_L$ and $\vR_T$, Eq.~(\ref{Boltz1}) 
can then be written as 
\begin{equation}
\frac{\partial \epsi_\nu}{\partial x_L}
\frac{\partial f_0}{\partial \epsi_\nu}
+ \frac{\partial \Delta f}{\partial x_L}
=  - \frac{\sigma_{A}}{V} \Delta f (x_L,R_T,\vk)~~ ,
\label{Boltz2}
\end{equation}
where $R_T \equiv |\vR_T|$ and  ${\partial \epsi_\nu}/{\partial
x_L} = ({\hat k} \cdot {\hat r}) {\partial \epsi_\nu}/{\partial
r}$.
The solution is given by
\begin{equation}
\Delta f(x_L,R_T,\vk) = \int_0^{x_L} d y
\left[ - \frac{\partial \epsi_\nu}{\partial y}
\frac{\partial f_0}{\partial \epsi_\nu} \right]
\exp\left[ - \int^{x_L}_{y} dz  \frac{\sigma_{A}}{V} \right] .
\label{BolSol}
\end{equation}

As neutrinos are created inside a PNS and propagate through the matter,  
their intensity will be  attenuated by  absorption. 
The exponential in Eq.~(\ref{BolSol}) accounts for this feature. 
If  $\sigma_{A}/{V}$ were sufficiently large, we would expect that 
very few neutrinos produced deep inside the PNS could
reach the surface. 
That, however, is not the case.

\subsubsection{Mean-Free-Path in NS matter}

\begin{figure}
\begin{center}
{\includegraphics[scale=0.48]{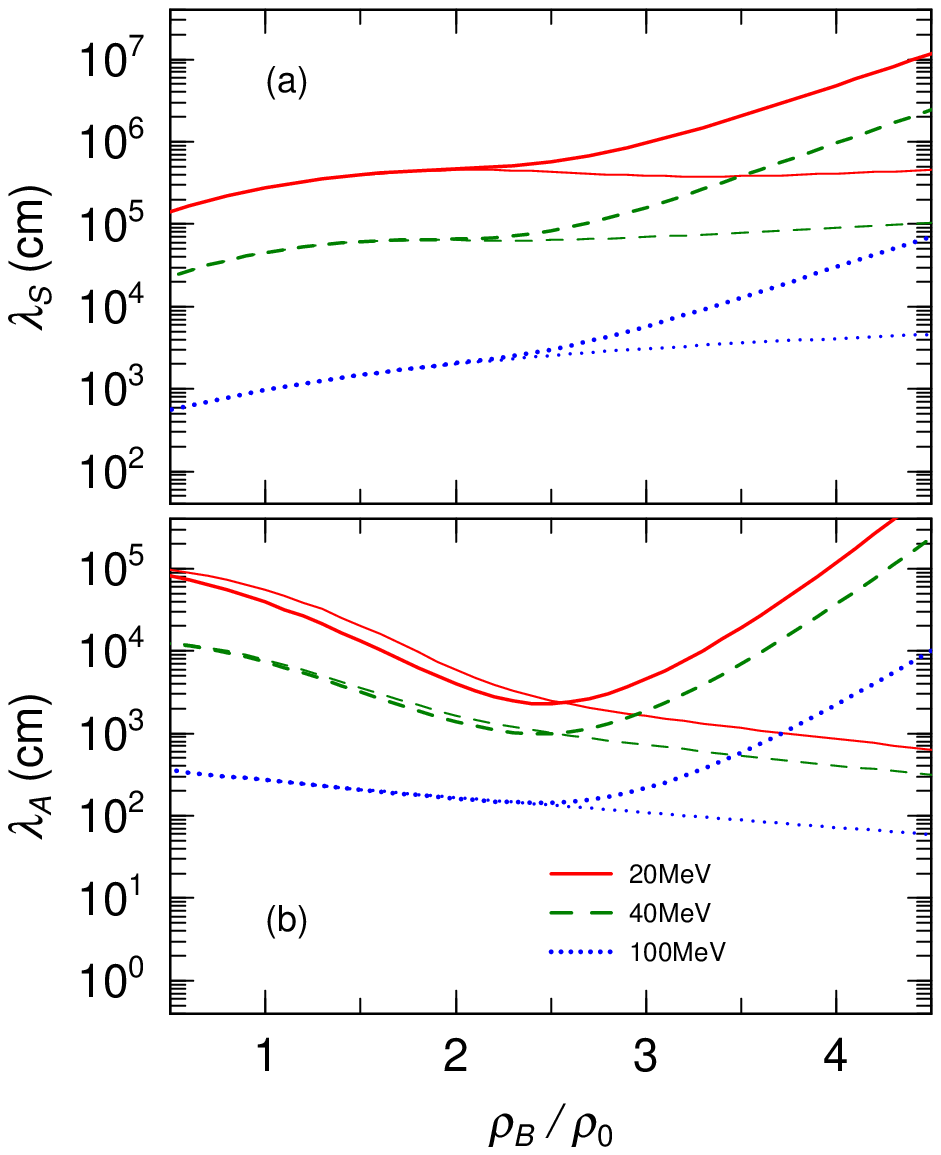}}
\caption{\small
(Color online) The neutrino mean-free-paths for
scattering (a) absorption (b) with a neutrino energy
$E_\nu = 20$ MeV (solid line),
40 MeV (dashed line) and 100 MeV (dotted line)
in  neutron-star matter at $T=20$ MeV without a magnetic field.
Thick and thin lines represent results with and
 without $\Lambda$s, respectively. }
\label{AbMgEn}
\end{center}
\end{figure}

To give a more concrete picture we next analyze  the mean-free path of
neutrinos. Fig.~\ref{AbMgEn} shows the neutrino MFPs for scattering and
absorption
$\lambda_{S,A} = (\sigma_{S,A}/V)^{-1}$
for   neutrino energies of $E_\nu = 20$ MeV (solid line),
$E_\nu = 40$ MeV (dashed line), and   $E_\nu = 100$ (dotted line)
in neutron-star matter at  a temperature of $T=20$ MeV without a magnetic field.

Thick and thin lines represent the results with and
 without  $\Lambda$s, respectively.
The MFPs for the absorption are less than a few km
so that most of the neutrinos produced in the central region are absorbed.
However, the neutrinos produced at the surface contribute to
the net emission of neutrinos;
this fact is qualitatively the same as the result obtained
in Ref.~\cite{arras99}.
In addition, we see that the neutrino MFP is longer when its energy is large
because of the Pauli blocking of the final electron.
As a result  lower energy neutrinos are absorbed more efficiently.

Furthermore, we should note that $\lambda_A \gg \lambda_s$ above nuclear matter density
 $\rho_B ~^>_\sim \rho_0$.
This highlights the fact that the absorption rate is much larger than
the scattering rate.  This is consistent to our approximation of
ignoring the scattering process.

In order to solve Eq.~(\ref{BolSol}), we need to know $\sigma_A/V$ as
a function of the density $\rho$, the magnetic field $B$, the
initial neutrino energy $E_{\nu}$, and the angle between the
magnetic field and the initial neutrino momentum, $\theta_{\nu}$. For
this calculation we have made a  data base of  $\sigma^0_{A}$ as a
function of the baryon density $\rho_B$ and the incident neutrino
energy $E_\nu$.

\begin{figure}
\begin{center}
{\includegraphics[angle=270, scale=0.5]{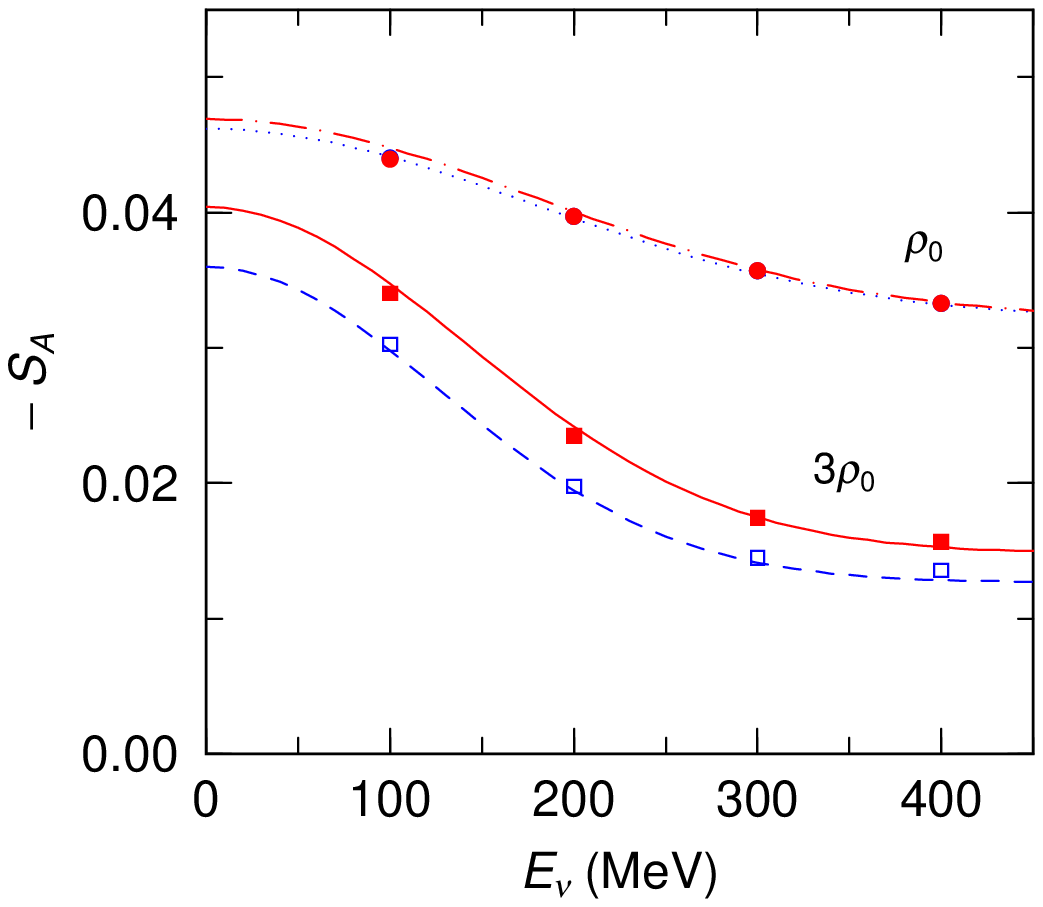}}
\caption{\small (Color online) The magnetic slope parameter Eq.~(67) of the neutrino
absorption  versus  incident energy  at $T=20$ MeV. Open and
full circles show the results in the present calculation at
$\rho_B=\rho_0$ without and with $\Lambda$s,
respectively. Open and full squares indicate those at
$\rho_B=3\rho_0$ without and with $\Lambda$s. Dotted,
dot-dashed, dashed and solid lines show results of the
fitting function at $\rho_B=\rho_0$ without and with
$\Lambda$s, and  those at $\rho_B=e\rho_0$ without and with $\Lambda$s,
respectively. }
\label{AbMgFt}
\end{center}
\end{figure}

However, it is not easy to make a data base of the magnetic
part of $\Delta \sigma_M$ because it is a function of  $\rho_B$, $E_\nu$ and
$\theta_\nu$ as well as $B$. This leads  to a computationally intensive  five dimensional
integration.
Therefore, we introduce a fitting function for the magnetic part
deduced  as follows.

\begin{table}[ht]
\caption{Parameters of Eqs.~(\ref{AM}) $-$ (\ref{CM}) fitted to theoretical results in Fig. 13}
\begin{tabular}{|c|c|c|}
\hline
& ~~ $p,n$ ~~ & ~~ $p,n,\Lambda$ ~~\\
\hline
$\cA_0$ & $7.28 \times 10^{-2}$ & $6.43 \times 10^{-2}$
\\ \hline
$\cA_1$ & $~4.07 \times 10^{-2}$ & $-3.22 \times 10^{-2}$
\\ \hline
$\gamma$ & 0.355 & 0.392
\\ \hline
$\cB_0$ & $2.96 \times 10^{-3}$ & $-2.62 \times 10^{-3}$
\\ \hline
$\cB_1$ & $7.21 \times 10^{-3}$& $2.36 \times 10^{-2}$
\\ \hline
$\cB_2$ & $5.94 \times 10^{-3}$& $-7.01 \times 10^{-3}$
\\ \hline
$\cB_3$ & $-2.02 \times 10^{-7}$&  $7.544 \times 10^{-4}$
\\ \hline
$\cC_0$ (MeV$^{2}$)& $ ~1.16 \times 10^{-5}$& $-1.05 \times 10^{-5}$
\\ \hline
$\cC_1$(MeV$^{2}$) & $2.29 \times 10^{-7}$& $-2.57 \times 10^{-6}$
\\ \hline
$\cC_2$ (MeV$^{2}$) & $-5.62 \times 10^{-6}$ & $-3.35 \times 10^{-6}$
\\ \hline
~$\cC_3$ (MeV$^{2}$)~ & $1.14 \times 10^{-6}$& $8.61 \times 10^{-7}$
\\ \hline
\end{tabular}
\label{PrTbl}
\end{table}

From Eq.~(\ref{SigAng}),
the angular dependence can be approximately written  as
\begin{equation}
\sigma_{A} = \sigma^0_A (1 + S_A \cos \theta_\nu) ,
\end{equation}
 where, $S_A$ obeys the following approximate function:
\begin{equation}
-S_A = \cA_M + \cB_M e^{-\cC_M E_\nu^2}
\label{AM}
\end{equation}
with
\begin{eqnarray}
\cA_M &=& \cA_0 + \cA_1 \left(\frac{\rho_B}{\rho_0}\right)^{\gamma}, \\
\cB_M &=& \cB_0 + \cB_1 \left(\frac{\rho_B}{\rho_0}\right)
+ \cB_2 \left(\frac{\rho_B}{\rho_0}\right)^2
+ \cB_3 \left(\frac{\rho_B}{\rho_0}\right)^3, \\
\cC_M &=& \cC_0 + \cC_1 \left(\frac{\rho_B}{\rho_0}\right)
+ \cC_2 \left(\frac{\rho_B}{\rho_0}\right)^2
+ \cC_3 \left(\frac{\rho_B}{\rho_0}\right)^3.
\label{CM}
\end{eqnarray}
All quantities except $\rho_B$ and $E_{\nu}$
are constant and adjusted to
reproduce the  theoretical results shown in Fig.~\ref{AbMgFt} as described in the figure caption.

\subsubsection{Proto Neutron-Star Model}

\begin{figure}[ht]
\begin{center}
{\includegraphics[scale=0.5,angle=270]{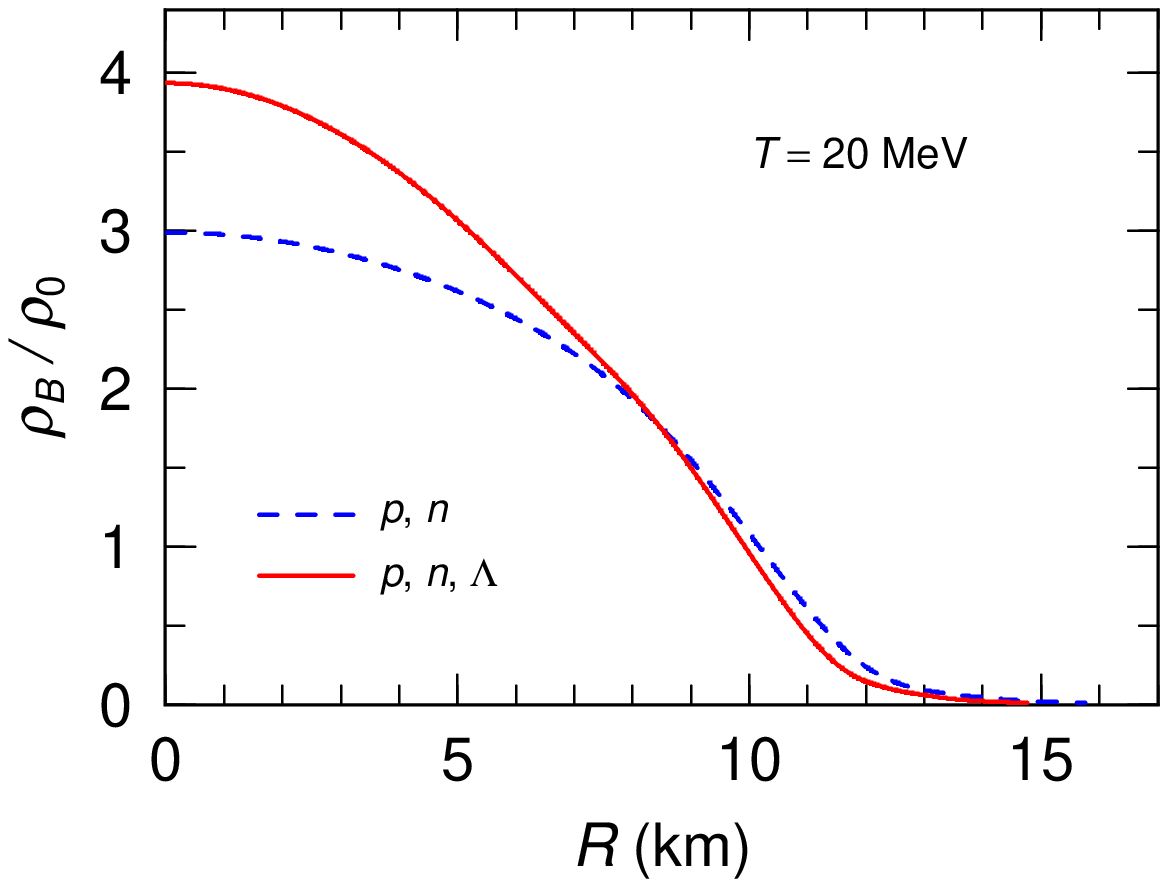}}
\caption{\small
(Color online) PNS Density distribution versus  radius.
Solid and dashed lines represent the results with
and without  $\Lambda$s at $T=20$ MeV, respectively.   }
\label{NtStar}
\end{center}
\end{figure}

To estimate the kick velocity in our model,
we need baryon density profiles of PNS.
Here, we assume an isothermal PNS mode which is easily calculated
and effective for our purpose in this work.
Baryon density profiles of our PNS model at $T=20$ MeV
are shown in Fig.~\ref{NtStar}.  
We choose 20 MeV as a reasonable average isothermal approximation to a PNS.  Even though the core temperature could be much more and the temperature at the neutrino sphere much less, 20 MeV is a reasonable average temperature encountered by neutrinos as they transport from the core to the neutrino-sphere.

For this illustration, we fix total gravitational mass of the PNS
to be 1.68 $M_{\odot}$.
The appearance of $\Lambda$ particles when $\rho_B\gtrsim 2\rho_0$
softens the EOS.
This increases the baryon density and the neutrino chemical potential.
The density profiles with $\Lambda$s are
sensitive to the temperature.

\subsubsection{Momentum transfer}

\begin{figure}[ht]
\begin{center}
{\includegraphics[scale=0.39,angle=270]{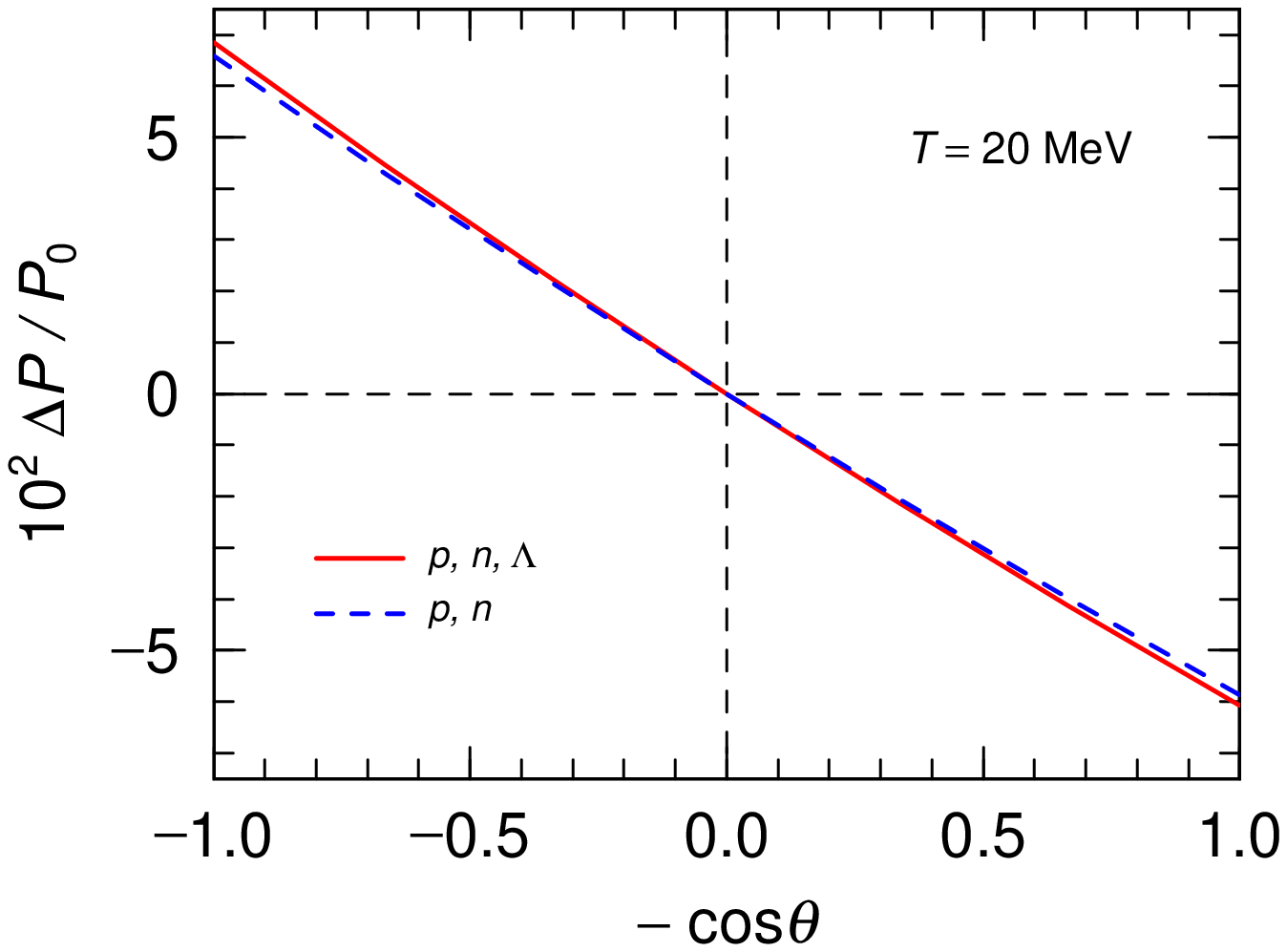}}
\caption{\small (Color online) The variation of emitted neutrino momentum versus
the polar direction.
Solid and dashed lines represent the results in a system with $\Lambda$s
and without $\Lambda$s at $T=20$ MeV, respectively. }
\label{FlAn}
\end{center}
\end{figure}

We use these density distributions of the PNS to the calculation of the neutrino momentum transport.
We define the effective spherical surface $S_N$ where
$\rho_B = \rho_0$,  and estimate the kick velocity from the
angular dependence of the emitted neutrino momentum at this surface.
The total momentum per unit time of the
neutrinos emitted along the direction $\vn$ is then calculated as
\begin{equation}
P =  \int_{S_N} d \vbr
\int \frac{d^3 k}{(2 \pi)^3} \Delta f (\vbr,\vk) ( \vk \cdot \vn )
\delta\left( \vk -  ( \vk \cdot \vn ) \vn \right)~~ .
\end{equation}
The momentum $P$ can be  approximately written as
\begin{equation}
P = P_0 + \Delta P \approx P_0 + P_1 \cos \theta ~~
\label{NtDis}
\end{equation}
in terms of  the polar angle $\theta$.
The asymmetry of the neutrino momentum $\Delta P / P_0$ is shown as a function of $\theta$ in Fig.~\ref{FlAn}.

We use  the momentum distribution in Eq.~(\ref{NtDis}) to calculate  the ratio of the
average momentum in the direction of the magnetic field $<P_z>$
to the total emitted neutrino energy $E_T$, {\it i.e.}
$<P_z>/E_T = P_1/ 3 P_0$.
Our results are estimated as
$<P_z>/E_T = 0.0194$ and 0.0176 with  and
 without the $\Lambda$s at $T= 20$ MeV.

We assume  that the  total  energy emitted in neutrinos is
$E_T \approx 3 \times 10^{53}$erg  \cite{lai98}.
For the   $M_{NS} = 1.68$ M$_{\odot}$ isothermal model with $T=20$ MeV, the calculated kick velocities are
$v_{kick} = <P_z> /M_{NS} = 580$~km s$^{-1}$ or  520~km s$^{-1}$
in neutron-stars with  or without $\Lambda$s, respectively.

In  actual observations,  the average value of the
kick velocity is about $v_{kick}$ = 400 km s$^{-1}$, and the highest reported value is $\sim 1500$ km s$^{-1}$.
Our values are thus close to the observed average pulsar kick velocity.
We note that Lai and Qian \cite{lai98} obtained  a similar result
($v_{kick}=$280 km $s^{-1}$).
However their result was calculated in a non-relativistic
framework without $\Lambda$ particles.

In the central region, high energy neutrinos up to $E_\nu \gtrsim 100$ MeV
are copiously  produced, but their MFP
is only about several $10^3$ cm.  They are, therefore,  almost completely
absorbed in the transport process.
The average energy of the emitted neutrino is about 20 MeV, and
most of neutrinos with energy $< 50$ MeV contribute to the pulsar kick
because the MFP for these neutrinos is larger (Fig.~\ref{AbMgEn}).
If one presumes that the thermalization process is faster than the time scale
at which the neutrino absorption process directly affects the collective
motion of the PNS, then the cross-section in the low density
region affects the final asymmetry of the neutrino emission.

Neutrinos are continuously further absorbed in the lower density regions before
they are emitted outside the neutron-star, and the asymmetry 
should be retained.
Indeed, when we  extend the  calculation to much lower density
$\rho_B = 0.5 \rho_0$,
we find that the asymmetries are almost the same as the above results, but that the
energy of the emitted neutrinos is small.

We caution, however, that in such  low density regions,  both the magnetic field and
temperature may be lower than those assumed in the present isothermal model.
If, instead of an isothermal neutron-star model, one were to use a an isoentropic model  with uniform entropy,
then the kick velocity may be smaller.
In the surface region of  magnetars, the magnetic field is still
as high as $10^{15}$ G. That is, however,  only about 1/100 of the value adopted in the present calculation.
A lower magnetic field may reduce $v_{kick}$, but
the lower density and temperature may enhance it.
In such a subtle situation the scattering process which we ignored in
the present calculation should also be included as it enhances
the neutrino asymmetry.  This could tend to  increase the kick velocity.

\newpage

\section{Summary and Discussion}

We have studied  the neutrino scattering and absorption processes
in strongly magnetized proto-neutron stars (PNSs) at finite temperature and density.
We used a  fully relativistic mean field (RMF) theory
for the hadronic sector of the EOS including hyperons.
We solved the Dirac equations for all constituent
particles,  p, n, $\Lambda$, e, and $\nu$, including a first order perturbation treatment
of a poloidal magnetic field with $B \sim 10^{17}$G.
We then applied  the solutions to obtain a  quantitative estimate
of the asymmetry that emerges from the neutrino-baryon collision processes.
We took  into account the Fermi motion of baryons and electrons, the momentum dependence of their spin vectors,  their recoil effects,
and the associated energy difference of the mean fields between the initial
and final particles exactly.
We thus included  the most important effects of the distortion of the Fermi spheres
made by the magnetic field in this fully microscopic framework, i.e. the asymmetric neutrino scattering and absorption cross-sections.

We found that the differential neutrino absorption cross-sections are
 suppressed
in the arctic direction parallel to the poloidal magnetic field $\vB$
in both cases with and without $\Lambda$s,
while the differential scattering cross-sections are slightly enhanced.
On the other hand, as expected from the sign of the couplings between the magnetic moments
of baryons and the external field,
the neutrino absorption and scattering cross-sections are respectively
enhanced and suppressed in the antarctic direction. This is completely opposite
to those in the arctic direction.
The differential cross-sections were integrated over the momenta
of the final electrons for absorption and over the momenta of initial neutrinos
for the scattering, respectively.
Quantitatively, when $B = 2 \times 10^{17}$G, the reduction for the absorption process
is  about 2\%, and the enhancement for the scattering process is about 1\%
in the forward direction along the direction of ${\vB}$.

Several interesting facts are evident  in the angular distributions
of  both cross-sections, which depend on the magnetic field, the baryon density,
and the temperature of the PNS matter.
Among them, we find, an appreciable forward suppression and
backward enhancement in the differential absorption cross-sections due to  the difference in Fermi distributions
between the spin-up and spin-down particles.  This effect is  larger at lower neutrino incident energy.
The asymmetry becomes smaller as the density  increases
because the asymmetry arises from the magnetic part of the cross-sections
which is proportional to the distortion of the Fermi surfaces
caused by the magnetic field.  This  tends to diminish with increasing matter density.

Using these cross-sections, we calculated the neutrino mean-free-paths (MFPs)
as a function of the baryon density and  temperature within a PNS.
We then applied  the above results to a calculation of  pulsar-kicks in core-collapse supernovae.
We solved the Boltzmann equation using  a one-dimensional
attenuation method, assuming that the neutrinos propagate along an approximately  straight line
and  that the system is in  steady state.
We  only included the MFPs for  neutrino absorption which dominates
over  scattering in producing the asymmetric momentum transfer to the PNS.

We estimated pulsar kick velocities from the calculated total momentum per unit time that  is transferred from the
emitted neutrinos to the PNS along the direction parallel to the poloidal magnetic field ${\vB}$.
For a 20-MeV isothermal neutron-star with $M_{NS} = 1.68 M_{\odot}$
and a total  energy in emitted neutrinos of $E_T \approx 3 \times 10^{53}$erg, the estimated kick velocities are
$v_{kick} = 580$~km s$^{-1}$ and  520~km s$^{-1}$ at $T=20$ MeV, including $\Lambda$s or no $\Lambda$s, respectively.
These values are in reasonable agreement with the observed average
pulsar-kick velocity of $v_{kick}$ = 400 km s$^{-1}$.

\section{Future Work}

In the present calculations we have adopted several assumptions
which we summarize here both as a caveat for the reader and
as a summary of issues to be addressed in future work.
One such assumption is  ignoring the neutrino scattering process in the solution of the Boltzmann equation.
This scattering  might enlarge the kick velocity.
The one-dimensional attenuation method to solve the Boltzmann equation
is also a coarse approximation.
We have  assumed that the asymmetry in neutrino emission is dominated
by the emission from  low-density regions with  $\rho_B \lesssim 3\rho_0$
where the neutrino opacity changes drastically.
We have also assumed that the internal high-density region only
contributes to the neutrino diffusive flux.  This diminishes  the expected neutrino asymmetry.
However, as was discussed in the last section,
if the thermalization process is considered dynamically
the asymmetric neutrino scattering and absorption in the high-density region
might also contribute to an aligned drift flux along the direction of
${\vB}$.
This could generate a  gradual acceleration of the pulsar-kick.
Numerical simulations of the neutrino transport inside a PNS
coupled to our microscopic calculations of the asymmetric
neutrino scattering and absorption cross-sections
are highly desirable in order to address to this critical question.
It has been  pointed out by Arras and Lai \cite{arras99}
that the neutrino distribution tends to be asymmetric
only near the surface of PNS.
This is consistent with the picture
adopted in the present attenuation approximation for the neutrino transport.
The issue becomes more subtle, however,
if the thermalization process is considered.
It would   be interesting to clarify by numerical calculations the  extent
to which the asymmetric
neutrino scattering and absorption contribute to the drift velocity
as well as the diffusive velocity of outgoing neutrinos considered here.
Other important questions are  to address the link among
asymmetric neutrino-baryon collisions, neutrino drift, and the collective
response of the PNS to the pulsar kick.
Further investigations must be done by numerically solving the Boltzmann
equation for the neutrino transport inside a PNS without approximations,
although we believe that our adopted scheme of attenuation
is more or less consistent with the microscopic picture.
Numerical calculations including several dynamical effects
are now underway.

We also should take account of  neutrino reactions
in the much lower density region, $\rho_B \gg \rho_0$, although we did not
include that in the present study because of the  numerical difficulty in calculating
thermodynamic quantities of the EOS in the RMF theory.
In such   low density regions, the magnetic field is weaker,
but the width of the Landau level, $\sqrt{2eB}$, could be of the same order
as the electron Fermi momentum, and it may affect the neutrino reactions.

The strength of the magnetic field inside the PNS  can easily
reach $3~-~4 \times 10^{18}$G in the high-density region according to
the scalar virial theorem.
This could make considerable effects, and a non-perturbative treatment of
the magnetic field must be applied for this high
 field strength \cite{Christian01}.
We may again need to take account of the Landau levels.

In this work we do not consider any magnetic contributions in the
neutrino production
\cite{chugai84,dorofeev,kisslinger2,kisslinger3,vosk86,parenkov}.
This also makes a contribution to the asymmetry of neutrino emissions.
As for the density profile of the PNS, we need to use an isoentropical model, in which the temperature becomes smaller in lower density region.
This effect may enhance the kick velocity.

We also did not take account of the resonant spin-flavor conversion
\cite{yoshida09} in the magnetized PNSs, and the neutrino-flavor conversion due to the MSW effect~\cite{kusenko96} or the self-interaction effect~\cite{duan06} in the present calculations.  All of these
could alter the asymmetric neutrino emission.
A quark-hadron phase transition~\cite{menezes09} or
a hyper-nuclear matter phase~\cite{rabhi10} under a strong magnetic
field is also  considered to be another source to affect the neutrino asymmetry.

If a poloidal magnetic field exists in the progenitor stars for SNe,
a stable toroidal magnetic field also is created in the core-collapse and explosion.
In this case, the angular dependence of the neutrino reactions may show a more
complicated and interesting behavior.
Thus, there are many open questions to be addressed in the future studies
which are beyond the scope of the present article.


\newpage

\appendix

\section{Dirac Spinor in a Magnetic Field}
\label{DSapp}

In this appendix, we explain the detailed expressions of the Dirac spinor under a magnetic field. The Dirac spinor $u(p)$ can be
obtained  by solving the following Dirac equation
\begin{equation}
{\hat K}(p) u (p,s) \equiv
\left[ \psla - M -  U_0 (b) -U_T \sigma_z \right ] u(p,s) =0 ,
\label{DiracEA}
\end{equation}
where $U_T = \mu B$. Here we defined the Green function $S(p)$ as
\begin{equation}
{\hat K}(p) S(p) = 1~~ .
\end{equation}
Then the Green function is written as
\begin{equation}
 S(p) =\det {\hat K} ( S_0 + S_1 U_T + S_2 U_T^2 + S_3 U_T^3 )~~,
\end{equation}
with
\begin{eqnarray}
\det {\hat K} &=& p_0^4 - 2 p_0^2 (\vp^2 + M^2 +U_T^2) + (\vp^2 + M^2)^2
+ 2 U_T^2 (p_z^2 - \vp_T^2 - M^2) +  U_T^4 ~~,
\nonumber\\
S_0 &= & (p_0^2 - E_p^2) (\psla + M) ~~,
\nonumber \\
S_1 &= &
(p_0^2 + E_p^2) \sigma_z + 2M p_0 \sigma_z  \gamma_0
-2 p_z (\vp \cdot \vsigma) \gamma_0
\nonumber \\ && \quad\quad\quad\quad\quad ~
+ 2M p_z \gamma_5 \gamma_0
+ 2i p_0 p_y \gamma^0 \gamma^1 - 2i p_0 p_x  \gamma^0 \gamma^2 ~~,
\nonumber \\
S_2 &= & - p_0 \gamma^0 + p_z \gamma^3 - p_x \gamma^1 - p_y \gamma^2 + M ~~,
\nonumber \\
S_3 &=& -\sigma_z ~~.
\label{SPrEl}
\end{eqnarray}

Here the single particle energy of this Dirac spinor
, which is obtained from $\det {\hat K} = 0$, becomes
\begin{equation}
e(\vp,s) = \sqrt{p_z^2 + \left( \sqrt{\vp_T^2 + M^2} + s U_T\right)^2}
= \sqrt{E_p^2 + 2 s U_T  \sqrt{\vp_T^2 + M^2} + U_T^2 } ~~,
\end{equation}
where $s=\pm1$, and $E_p = \sqrt{\vp^2 + M^2}$.
Then, $\det {\hat K}$ is rewritten as
\begin{equation}
\det {\hat K} = (p^2_0 - e^2(\vp,1)) (p^2_0 - e^2(\vp,-1))~~ .
\end{equation}

Furthermore, the Green function for  this particle is written as
\begin{eqnarray}
S(p) &=& {\hat K}^{-1}(p) =
\sum_{s=\pm 1} \frac{u(\vp,s) {\bar u}(\vp,s)}{p_0 - e(\vp,s) \pm i \delta}
+\sum_{s=\pm 1} \frac{v(-\vp,s) {\bar v}(-\vp,s)}{p_0 + e(\vp,s) + i \delta} ~~,
\end{eqnarray}
where $u(\vp,s)$ and $v(-\vp,s)$ are the Dirac spinors of the
positive and negative energy states, respectively.

By using the above quantities, we can obtain the Dirac spinor as
\begin{equation}
u(\vp,s) {\hat u}(\vp,s) =  \lim_{p_0 \to e({\bf p},s)}(p_0 - e(\vp,s)) S(p)~~.
\end{equation}

Now we expand $S$ with respect to $U_T$ and determine the Dirac spinor
in  first order perturbation theory.
Here we define
\begin{eqnarray}
D_e &\equiv & \lim_{p_0 \to e({\bf p},s)} \frac{p_0 - e(\vp,s)}{\det {\hat K}}
 = \frac{1}{ 8 e(\vp, s) \left( s U_T \sqrt{\vp_T^2 + M^2}  \right) }~~.
\end{eqnarray}

When $|U_T| \ll 1$, we can substitute
$p_0 = e(\vp,s) \approx E_p + sU_T \sqrt{\vp_T^2 + M^2}/E_p$
into Eq.~(\ref{SPrEl}) and obtain
\begin{eqnarray}
 D_e S_0 & \approx &
 \frac{s}{ 4 E_p }
\left(  1 - \frac{s U_T \sqrt{\vp_T^2 + M^2} }{E_p^2} \right)
\left( 1 + \frac{s U_T}{2 \sqrt{\vp_T^2 + M^2}}  \right)
\nonumber \\ && \quad\quad\quad \times
\left\{ (\psla+M) +
\frac{sU_T \sqrt{\vp_T^2 + M^2}}{E_p} \gamma_0 \right\}
\nonumber \\ & \approx & \frac{1}{4E_p}
\left\{ (\psla+M) + \left[
\frac{\sqrt{\vp_T^2 + M^2}}{E_p} \gamma_0
\right. \right.
\nonumber \\ && \quad\quad\quad\quad\quad\quad\quad\quad
\left.\left.
+ \frac{p_z^2 - \vp_T^2 - M^2 }{2 E_p^2\sqrt{\vp_T^2 + M^2}}
(\psla+M) \right] s U_T \right\}_{p_0 = E_p}
\\
U_T D_e S_1 & \approx &
 \frac{s}{ 8 E_p \sqrt{\vp_T^2 + M^2}}
\left(  1 - \frac{s U_T \sqrt{\vp_T^2 + M^2} }{E_p^2} \right)
\nonumber \\ & \times &
 \left\{S_1 + 2 ( E_p \sigma_z + M \gamma_0 \sigma_z
+ i p_y \alpha_x - i p_x \alpha_y
) \frac{s \sqrt{\vp_T^2 + M^2}}{E_p} U_T \right\}_{p_0=E_p}
\nonumber \\
& \approx &
 \frac{1}{ 4 E_p }
\left\{ \frac{S_1}{ \sqrt{\vp_T^2 + M^2} }
\right. \nonumber \\ &&   \left. \quad\quad
+ ~ U_T \left[ -  \frac{S_1}{2 E_p^2}
+ \frac{1}{E_p}( E_p \sigma_z + M \gamma_0 \sigma_z + i p_x \sigma_x
- i p_y \sigma_y ) \right] \right\}_{p_0 =  E_p}
\nonumber \\
& \approx &
\frac{1}{ 4 E_p }
\left\{ s (\psla + M)\gamma_5 \asla +  \frac{p_z}{E^2_p}
\left( \beta {\vsigma} \cdot \vp - M \gamma_5 \right) U_T \right\}_{p_0 =  E_p}
\nonumber \\
U_T^2 D_e S_2 & \approx &
\frac{s U_T}{ 8 E_p \sqrt{\vp_T^2 + M^2} }
\left(- E_p \gamma^0 + M + p_z \gamma^3 - p_x \gamma^1 - p_y \gamma^2
\right)~~,
\end{eqnarray}
with
\begin{equation}
a =  \frac{1}{\sqrt{\vp_T^2 + M^{2}}}( p_z, 0, 0, E_p) ~~.
\end{equation}

Then, the Dirac spinor is written as up to the first order in  $U_T$
\begin{eqnarray}
u(\vp,s){\bar u}(\vp,s) & \approx &
 \frac{(\psla + M)(1 + \gamma_5 \asla(p)s )}{4 E_p}
+  \frac{p_z U_T}{ 4 E^3_p}
\left( {\vsigma} \cdot \vp - M \gamma_5 \gamma_0 \right)
\nonumber \\ && \quad\quad
+ \frac{s U_T}{ 8 E_p \sqrt{\vp_T^2 + M^2} }
\left(- E_p \gamma^0 + M + p_z \gamma^3 - p_x \gamma^1 - p_y \gamma^2
\right)~~.
\end{eqnarray}

\section{Neutrino Reaction Cross-Sections}
\label{NRCS}

In this appendix, we derive Eqs.~(\ref{DsigM1}) and (\ref{DsigEL}). 
We start from the product of leptonic and hadronic weak currents 
in Eq.~(\ref{WBL}). 
By considering the spin-dependence, we express the $W_{BL}$ 
in Eq.~(\ref{WBL}) as follows
\begin{equation}
W_{BL} = W_0 + W_i s_i + W_f s_f + W_e s_l + W_2 s_i s_f + W_3 s_l s_i +
 W_4 s_l s_f .
\label{B-WEL}
\end{equation}
Note that $W_e$, $W_3$ and $W_4$ only appear  when the final lepton is 
an electron.

When $|\mu_b B| \ll \epsi_b - U_0(b)$, the baryon Fermi distribution
function can be  expanded as
\begin{equation}
n_b(e_b(\vp,s)) \approx n_b (E_b^*(\vp) + U_0(b))
+ n^{\prime}_b(E_b^*(\vp)+U_0(b))\Delta E_b(\vp)  s ,
\label{B-Bex}
\end{equation}
and the electron distribution is written as
\begin{equation}
n_e(e_e (\vk)) \approx n_e(|\vk|)
+  n^{\prime}_e(\vk) \frac{m_e}{|\vk|} \mu_e B s_l ,
\label{Eex}
\end{equation}
where $n^\prime_b (x) = \partial n_b(x)/ \partial x$.
In addition, the energy delta-function in Eq.~(\ref{CrsLB}) is also expanded as
\begin{eqnarray}
&&\delta(|\vk_i|+ e_i(\vp_i,s_i) - e_l(\vk_f, s_l) - e_f(\vp_f,s_f))
\nonumber \\ &\approx&
\delta(|\vk_i|+ E_{\alpha}^{*}(\vp_i) + U_0(\alpha) - |\vk_f| - E_{\beta}^{*}(\vp_f) - U_0(\beta))
\nonumber \\ &&
+ \delta^\prime(|\vk_i|+ E_{\alpha}^{*}(\vp_i) + U_0(\alpha) - |\vk_f| - E_{\beta}^{*}(\vp_f) - U_0(\beta))
\Delta E~~ ,
\end{eqnarray}
where $\delta^\prime (x) \equiv  \partial \delta (x)/ \partial x$,
and
\begin{equation}
\Delta E = 
\Delta E_{\alpha} (\vp_i) s_i - \Delta E_{\beta}(\vp_f) s_f
- \frac{m_e}{|\vk|} \mu_e B s_l \delta_{l, e} ~~.
\end{equation}
Here, we define the momentum transfer $q = (q_0,\vq)$ as
\begin{equation}
q \equiv (|\vk_i| - |\vk_f|  - \Delta U_0; \vk_i - \vk_f)
\end{equation}
with $\Delta U_0 = U_0(\beta)-U_0(\alpha)$, and rewrite the energy
delta-function as
\begin{eqnarray}
&&\delta(|\vk_i|+ E_{\alpha}^{*}(\vp_i) + U_0(\alpha) - |\vk_f| - E_f^{*}(\vp_i+\vq) - U_0(\beta))
\nonumber \\ &=&
\delta(E_{\alpha}^{*}(\vp_i) + q_0 - E_{\beta}^{*}(\vp_i + \vq))
= \frac{E_{\beta}^*}{|\vp_i||\vq|} \delta(t - t_p)~~,
\end{eqnarray}
where $t \equiv \vq \cdot \vp_i/(|\vq| |\vp_i|)$, and
\begin{equation}
t_p = \frac{2 q_0 E_{\alpha}^*(\vp_i) + q^2 + M_i^{*2} -  M_{\beta}^{*2}}{2|\vq||\vp_i|} ~~.
\end{equation}
Furthermore we write
\begin{equation}
\delta^\prime(E_{\alpha}^{*}(\vp_i) + q_0 - E_{\beta}^{*}(\vp_i + \vq))
= \frac{1}{|\vp_i||\vq|} \delta(t - t_p)
+  \frac{E_{\beta}^{* 2}}{\vp_i^2 \vq^2} \frac{\partial}{\partial
t}\delta(t - t_p) ~~.
\end{equation}

Note that the terms proportional to $s_{\kappa}$ ($\kappa = l,i,j$)
vanish in Eq. (B5), and the $W_{2,3,4}$ do not
contribute to the final results to first order in $\mu_b B$.
In view of  this fact, we can further separate the magnetic part of the
cross-section of Eq.~(\ref{CrsNM}) into  two parts as
\begin{equation}
\Delta \sigma = \Delta \sigma_M + \Delta \sigma_{el},
\label{B-Dsig}
\end{equation}
where the first and second terms are the contributions from the target
particle and the outgoing electron, which appear only in the
absorption ($\nu_e \rightarrow e^{-}$) process. Detailed expressions of each term are presented at the Eqs.(\ref{DsigM1}) $\sim$ (\ref{Wel}) in text.

\acknowledgments
This work was supported by the Grants-in-Aid for the Scientific
Research from the Ministry of Education, Science and Culture of
Japan~(20244035, 21540412),  Scientific Research on Innovative Area
of MEXT (20105004), and Heiwa Nakajima Foundation.
This work was partially supported by the National Research Foundation
of Korea (Grant No. 2011-0003188, 2011-0015467).
Work at the University of Notre Dame (G.J.M.) supported
by the U.S. Department of Energy under
Nuclear Theory Grant DE-FG02-95-ER40934.
%



\end{document}